\documentclass[amsmath,amssymb,prx,aps,showpacs,superscriptaddress,twocolumn,10pt]{revtex4-1}
\usepackage{amsmath,amsfonts,amssymb,amsthm,graphics,graphicx,epsfig,bbm}
\usepackage[colorlinks=true,citecolor=blue,linkcolor=red,urlcolor=blue]{hyperref}
\usepackage[usenames]{color}
\usepackage{graphicx}
\usepackage{subfigure}
\usepackage{amsmath}
\usepackage{epsfig}
\usepackage{dcolumn}
\usepackage{bm}
\usepackage{color}
\usepackage{epstopdf}
\usepackage{amssymb}
\usepackage{amstext}
\usepackage{latexsym}
\usepackage{hyperref}
\usepackage{amsfonts}
\usepackage{psfrag}
\usepackage{xcolor}
\usepackage[normalem]{ulem}
\usepackage{dsfont}
\usepackage{txfonts}
\usepackage{overpic}

\renewcommand{\Re}{\mathrm{Re}}

\newcommand{\ket}[1]{\vert #1 \rangle}
\newcommand{\bra}[1]{\langle #1 \vert}

\newcommand{\abs}[1]{| #1 |}

\begin{document}

%%%%%%%%%%%%%%%%%%%%%%%%%%%%%%%%%%
\title{
 Fully-programmable universal quantum simulator with a one-dimensional quantum processor}
%%%%%%%%%%%%%%
%%%%%%%%%%%%%%%%%%%%
\author{V. M. Bastidas}
\altaffiliation[]{These authors contributed equally to this work.}
%\email{victor.m.bastidas.v.yr@hco.ntt.co.jp }
\affiliation{NTT Research Center for Theoretical Quantum Physics, NTT Corporation, 3-1 Morinosato-Wakamiya, Atsugi, Kanagawa, 243-0198, Japan}

\author{T. Haug}
\altaffiliation[]{These authors contributed equally to this work.}
\affiliation{NTT Research Center for Theoretical Quantum Physics, NTT Corporation,  3-1 Morinosato-Wakamiya, Atsugi, Kanagawa, 243-0198, Japan} 
\affiliation{Centre for Quantum Technologies, National University of Singapore, 3 Science Drive 2, Singapore 117543, Singapore}

\author{C. Gravel}
\altaffiliation[]{Present address: EAGLYS, \#301 Utsumi Building, 1-55-14 Yoyogi, Shibuya-ku, Tokyo, 151-0053.}
\affiliation{National Institute of Informatics, 2-1-2 Hitotsubashi, Chiyoda-ku, Tokyo 101-8430, Japan}

\author{L.-C. Kwek}
\affiliation{Centre for Quantum Technologies, National University of Singapore, 3 Science Drive 2, Singapore 117543, Singapore%
}

\affiliation{MajuLab, CNRS-UNS-NUS-NTU International Joint Research Unit, UMI 3654, Singapore
}

\affiliation{Institute of Advanced Studies, Nanyang Technological University, 60 Nanyang View, Singapore 639673, Singapore%
}

\affiliation{National Institute of Education, Nanyang Technological University, 1 Nanyang Walk, Singapore 637616, Singapore%
}

\author{W. J. Munro}
\affiliation{NTT Research Center for Theoretical Quantum Physics, NTT Corporation, 3-1 Morinosato-Wakamiya, Atsugi, Kanagawa, 243-0198, Japan}
\affiliation{National Institute of Informatics, 2-1-2 Hitotsubashi, Chiyoda-ku, Tokyo 101-8430, Japan}

\author{Kae Nemoto}
\email{nemoto@nii.ac.jp }
\affiliation{National Institute of Informatics, 2-1-2 Hitotsubashi, Chiyoda-ku, Tokyo 101-8430, Japan}
%%%%%%%%%%%%%%%%%%%%%%%%%%%%%%%%%

\date{\today}

%%%%%%%%%%%%%%%%%%%%%%%%%%%%%%%%%

\date{\today}

\begin{abstract}
Current quantum devices execute specific tasks that are hard for classical computers and have the potential to solve problems such as quantum simulation of material science and chemistry, even without error correction. For practical applications it is highly desirable to reconfigure the connectivity of the device, which for superconducting quantum processors is determined at fabrication. In addition, we require a careful design of control lines and couplings to resonators for measurements. Therefore, it is a cumbersome and slow undertaking to fabricate a new device for each problem we want to solve. Here we periodically drive a one-dimensional chain to engineer effective Hamiltonians that simulate arbitrary connectivities. We demonstrate the capability of our method by engineering driving sequences to simulate star, all-to-all, and ring connectivities. We also simulate a minimal example of the 3-SAT problem including three-body interactions, which are difficult to realize experimentally.
Our results open a new paradigm to perform quantum simulation in near term quantum devices by enabling us to stroboscopically simulate arbitrary Hamiltonians with a single device and optimized driving sequences
\end{abstract}

\maketitle
%%%%%%%%%%%%%%%%%%%%%%%%%%%%%%%%%

%%%
\begin{figure*}
 \centering
 \includegraphics[width=0.90\textwidth]{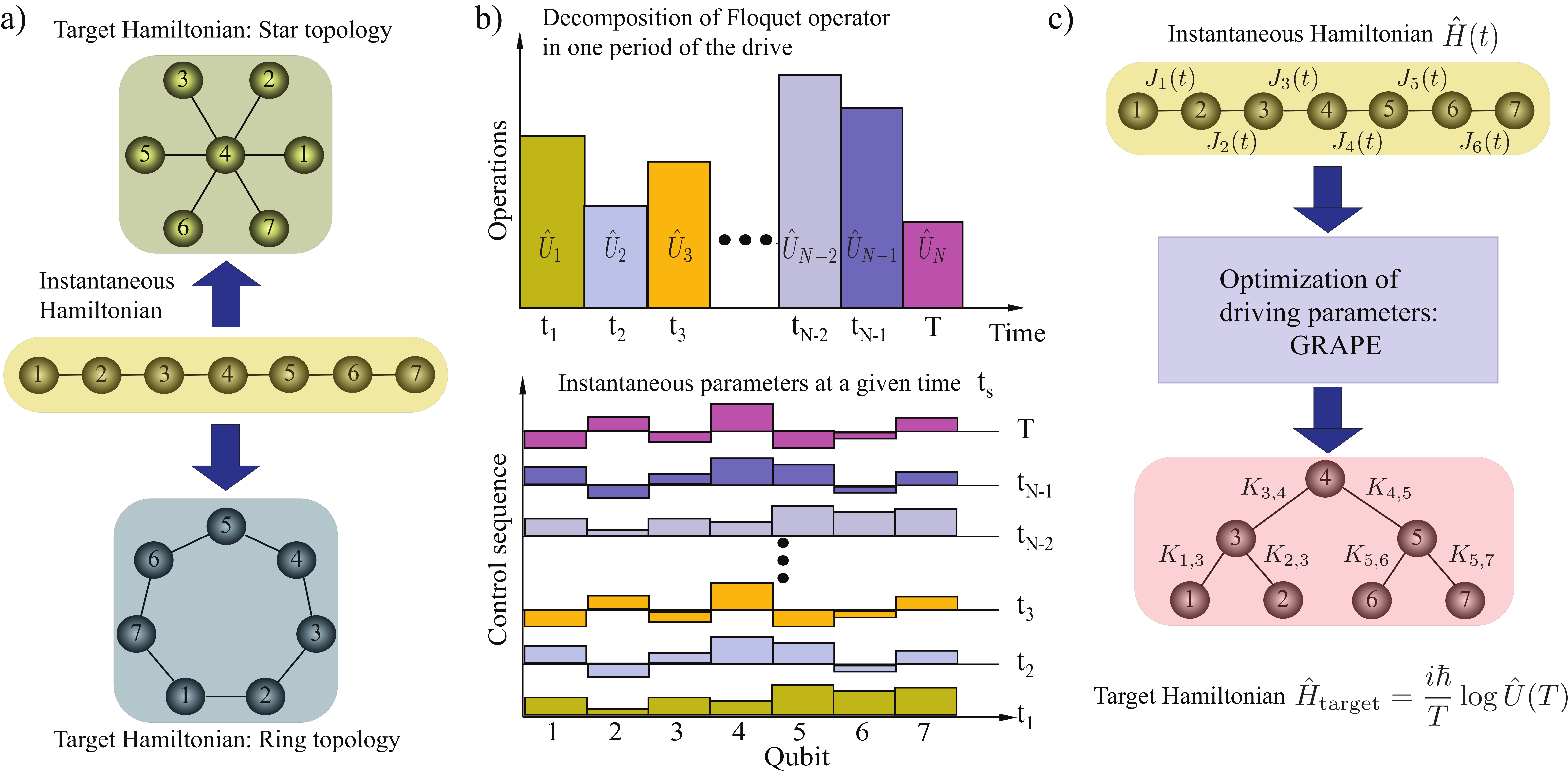}
	\caption{Generating effective topologies using Floquet engineering and by optimizing the driving sequence. a) Schematic diagram illustrating our Hamiltonian engineering approach. Here we consider a device with a simple topology at each time $t$, described by the instantaneous Hamiltonian $\hat{H}(t)=\hat{H}(t+T)$, where $T$ is the period of the external control. The latter allows to simulate other connectivities at stroboscopic times $t=nT$ that are difficult to achieve.
	b) Depicts the decomposition of the unitary evolution within a period of the drive as a product of unitaries $\hat{U}_j$ in the time interval $t_{j-1}\leq t\leq t_{j}$ with $j\in\{1,\ldots,N\}$ and the driving sequence applied to the qubits at each time $t_j$. c) Considers a driven quantum device, described by the instantaneous Hamiltonian $\hat{H}(t)$. By optimizing over the different driving sequences, one can determine how to drive the system in order to achieve the goal of producing the target Hamiltonian $\hat{H}_{\text{target}}$ with the desired topology.
 }
\label{Fig1}
\end{figure*}
 %%%

%%%%%
\section{Introduction}
\label{SecI}
%%%%%
The dynamics of driven many-body systems has many relevant applications in quantum technologies and quantum information processing~\cite{Amico2008,cheneau2012light,Nemoto2014,lewis2019dynamics}. Existing quantum devices allow us to perform tasks such as quantum metrology~\cite{Degen2017}, quantum communication~\cite{Hasegawa2019} and quantum simulation~\cite{roushan17, 2019Schuster,Yangsen2019,yan2019strongly,Roushan2017chiral,chiaro2019,o2016scalable,kandala2017hardware,arute2020hartreefock}, that can be carried out without error correction. Among the Noisy Intermediate-scale Quantum (NISQ) devices, superconducting~\cite{roushan17, 2019Schuster,Yangsen2019} and trapped ions~\cite{lanyon2011universal,blatt2012quantum, zhang2017observation} quantum processors play a fundamental role and have huge potential for near-term applications~\cite{preskill2018quantum}. These devices are highly programable and potentially able to outperform classical computers on specific tasks~\cite{arute2019quantum}. The available processors can be controlled locally through control lines, enabling the tuning of parameters of the Hamiltonian such as on-site energies and couplings~\cite{Roushan2017chiral,chiaro2019,Yangsen2019,yan2019strongly,zha2020ergodic}. Therefore, given the high degree of control the current technology already allows, it is now necessary to design a pathway towards practical applications of such devices.

Quantum machines promise dramatic speed-up in computation time for the simulation of quantum chemistry and many-body problems~\cite{Babbush2018,georgescu2014quantum}. However, a major challenge lies in encoding the actual problem onto the limited physical connectivity of the quantum simulator. Ideally, a quantum simulator would be fully-programmable such that it can achieve direct connections between arbitrary qubits that can be modified at any time.
Unfortunately, each device has a fixed and limited connectivity and it is quite challenging to reconfigure it after fabrication, such that it is able to perform another task.  In the long term, a fault tolerant quantum computer would allow this but that option is not available in the NISQ regime\cite{preskill2018quantum}. In practice, due to the hardware restrictions,  one simply designs and builds quantum processors tailored to solve the particular problem of interest. This current approach to quantum simulators is quite limited and time-consuming.

In this work, we enhance NISQ devices with tailored driving sequences to become fully re-programmable quantum simulators.  Our approach is based on modest assumptions of local controllability of a quantum device, which are already available in many systems~\cite{Roushan2017chiral,chiaro2019,Yangsen2019,yan2019strongly,zha2020ergodic,Wei2019,Rubio2020}. This allows one to apply a periodic drive to control different parameters of the system. Thus, the 
Hamiltonian is periodic in time with $\hat{H}(t)=\hat{H}(t+T)$. Associated with this is the evolution operator within a period of the drive $\hat{\mathcal{F}}=\hat{U}(T;0)$, known as the Floquet operator~~\cite{bastidas2018ergodic,RevModPhys.89.011004,PhysRevLett.117.250401,bukov2015universal,oka2019floquet}.
As the Floquet operator is unitary, it is natural to define an effective Hamiltonian such that $\hat{\mathcal{F}}=\exp(-\mathrm{i}\hat{H}_{\text{eff}}T/\hbar)$. Now if we measure the dynamics of the system at stroboscopic times $t=nT$, the device described by the instantaneous Hamiltonian $\hat{H}(t)$ can be used to simulate the effective Hamiltonian $\hat{H}_{\text{eff}}$. Within the framework of Floquet engineering, the usual approach is that for a given driving, one can
analytically obtain the effective Hamiltonian in the high-frequency regime~\cite{RevModPhys.89.011004,PhysRevLett.117.250401,bukov2015universal,oka2019floquet}.  However, the inverse problem is highly nontrivial, because it requires the design of an appropriate driving sequence to the device in order to achieve a desired target Hamiltonian $\hat{H}_{\text{target}}$.
 Here, we provide a solution to this problem using GRAPE, an optimization algorithm that allows us to obtain the desired driving sequence. In figure~\ref{Fig1}~a) we depict an example of our idea, where we drive a linear chain to simulate star and ring topologies.  A related work~\cite{Onodera2020quantum}, focuses on stroboscopic simulation of effective Hamiltonians by driving a device with all-to-all coupling to achieve a desired effective Hamiltonian. This approach, however, requires coupling to an external bus, which can propagate errors to the whole system.

Our paper is organized as follows. In Sec.~\ref{SecII} we introduce the Bose Hubbard Hamiltonian describing interacting microwave photons and discuss the hardcore boson limit. In Sec.~\ref{SecIII} we discuss GRAPE, an optimization algorithm used to find an optimal driving sequence to simulate the target Hamiltonian. We also discuss the family of Hamiltonians that can be simulated using our method. In Sec.~\ref{SecIV} we show how to apply our method to simulate arbitrary topologies by driving a one dimensional quantum processor. After that, in Sec.~\ref{SecV} we discuss how to apply our method to stroboscopically simulate a 3-Sat problem and its adiabatic deformation. We also show how to use our scheme to simulate problems in quantum chemistry in Sec.~\ref{SecVI}. 

%%%%%
\section{Model}
\label{SecII}
%%%%%
Here we tailor our approach to one of the most promising NISQ devices based on a one-dimensional array of $L$ coupled superconducting qubits, described by the Bose-Hubbard Hamiltonian~\cite{roushan17,chiaro2019,Yangsen2019,yan2019strongly,zha2020ergodic}
%%%
\begin{align}
\label{eq:HamiltonianResonatorArray}
\hat{H}(t)&=\hbar\sum^{L}_{l=1}\left[g_{l}(t) \hat{n}_{l}+\frac{U}{2} \hat{n}_{l}(\hat{n}_{l}-1)\right]+\hbar \sum^{L-1}_{l=1}J_l(t)(\hat{a}^{\dagger}_{l}\hat{a}_{l+1}+\text{h.c})
 \ ,
 \end{align}
%%%
 where $\hat{n}_{l}=\hat{a}^{\dagger}_{l}\hat{a}_{l}$ is the number operator, $\hat{a}_{l}$ and $\hat{a}^{\dagger}_{l}$ are bosonic annihilation and creation operators at site $l$, respectively.
With current experimental feasibilities, it is possible to use Z and XY control lines of the superconducting processor to drive the angular frequencies of the qubits $g_{l}(t)$ and the coupling strengths $J_l(t)$.  However, the anharmonicity $U$ is kept fixed and it is determined from fabrication. The latter plays an important role in the dynamics. For example, in the hardcore-boson regime $U\gg g_{l}, J_{l}$, the large anharmonicity prevents two microwave photons from being at the same site~\cite{Yangsen2019,yan2019strongly}. In this case, the Hamiltonian~\eqref{eq:HamiltonianResonatorArray} can be written in terms of Pauli matrices $\sigma^{\alpha}_l$ with $\alpha\in\{x,y,z\}$, as follows
 %%%
\begin{align}
\label{eq:HamiltonianHarcoreBoson}
\hat{H}(t)&=\frac{\hbar}{2}\sum^{L}_{l=1}g_{l}(t) \sigma^z_{l}+\frac{\hbar}{2} \sum^{L-1}_{l=1}J_l(t)(\sigma^x_{l}\sigma^x_{l+1}+\sigma^y_{l}\sigma^y_{l+1})
 \ .
 \end{align}
%%%
 It is useful to mention that the Hamiltonians~\eqref{eq:HamiltonianResonatorArray}~and~\eqref{eq:HamiltonianHarcoreBoson}  preserve the total number $M$ of excitations. By repeating a control sequence in a periodic fashion with a period $T$, the Hamiltonians~\eqref{eq:HamiltonianResonatorArray}~and~\eqref{eq:HamiltonianHarcoreBoson} become periodic in time, i.e., $\hat{H}(t)=\hat{H}(t+T)$. Therefore, we can use Floquet theory to explore the dynamics of the system. 

The cornerstone of our approach is
 to use an optimization algorithm to determine a driving sequence that is applied to a one-dimensional quantum processor. To achieve that, we consider the decomposition of the Floquet operator as a product of $N$ unitary operators $\hat{U}_j$
%%%
\begin{equation}
\label{eq:FloquetDecomposition}
        \hat{\mathcal{F}}=\hat{U}(T;0)=\hat{U}_N\hat{U}_{N-1}\cdots\hat{U}_{2}\hat{U}_{1}
\end{equation}
%%%
 at different time steps $t_j=j \tau$ with $\tau=T/N$.
This decomposition can be obtained by keeping the parameters of the system constant in certain time intervals. We can fix a set of values of the frequencies $g_{l}(t)=\bar{g}^j_l$ or couplings $J_{l}(t)=\bar{J}^j_l$  for all the qubits in the time interval $t_{j-1}\leq t\leq t_{j}$. If we let the system evolve during that time interval, we obtain the evolution operator $\hat{U}_j=\exp(-\mathrm{i}\hat{H}(t_j)\tau/\hbar)$. The basic idea of the control scheme is depicted in Fig.~\ref{Fig1}~b). Given a control sequence with $N$ time steps, it is not trivial a priori to know which values $\bar{g}^j_l$ and $\bar{J}^j_l$
we should choose to obtain our desired target Hamiltonian.

%%%%%%%%%%%%%%%%%%%%%%%%%%%%%%%%%%%%%%%
\section{Optimization algorithm to find the driving sequence and effective Hamiltonians}
\label{SecIII}
Here we translate the task of finding the driving sequence into an optimization problem that we solve using GRAPE~\cite{machnes2011comparing,johansson2013qutip,Wu2019}, which is a well-known optimization algorithm. By using this method, we can simulate effective Hamiltonians for finite $U$ and in the hardcore boson regime, as we show below.
%%%%%
\subsection{GRAPE: Optimization algorithm}
%%%%%
The goal is to vary the controls in time, such that a specific target unitary $\hat{\mathcal{F}}_{\text{target}}=\hat{U}_{\text{target}}(T)=\exp(-\mathrm{i}\hat{H}_{\text{target}}T/\hbar)$ is reached. The equation of motion for the unitary is given by
\begin{equation}
i\hbar\frac{d}{dt}\hat{U}(t)=\left(\hat{H}_\text{d}+\sum_{k=1}^R u_k(t)\hat{V}_k\right)\hat{U}(t) \, ,
\end{equation}
with a constant drift Hamiltonian term $\hat{H}_\text{d}$ and the $R$ control terms $\hat{V}_j$, which vary with $u_j(t)$ in time. In this section, $ \hat{H}_u(t)=\hat{H}_\text{d}+\sum_{k=1}^R u_k(t)\hat{V}_k$ denotes the control Hamiltonian associated with a control sequence $\{u_k(t)\}$. They correspond to the driving sequence $\bar{g}^j_l$ and $\bar{J}^j_l$ in the Hamiltonians~\eqref{eq:HamiltonianResonatorArray}~and~\eqref{eq:HamiltonianHarcoreBoson}.
For each set of control parameters, we can produce a trial Floquet operator $\hat{\mathcal{F}}_{\text{trial}}=\hat{U}_{\text{trial}}(T)$.
To measure how well the control is achieving the task, we determine the fidelity defined by
%%%
\begin{equation}
F=\frac{1}{D_{M,L}}\abs{\text{tr}(\hat{\mathcal{F}}_{\text{target}}^\dag \hat{\mathcal{F}}_{\text{trial}})} \, ,
\end{equation}
%%%
where $D_{M,L}$ is the Hilbert space dimension for $M$ bosons in $L$ sites  and $T$ is the drive period. The value of $F$ can take values between 0 and 1, where 1 is reached when the $\hat{\mathcal{F}}_{\text{trial}}$ corresponds to $\hat{\mathcal{F}}_{\text{target}}$.
The optimal $\hat{\mathcal{F}}_{\text{trial}}$ with $F=1$ are not unique, as there can be global phases in the unitary that do not affect the overall dynamics and only introduce a phase in $\text{tr}(\hat{\mathcal{F}}_{\text{target}}^\dag \hat{\mathcal{F}}_{\text{trial}})$. However, these global phases can change the effective Hamiltonian found by $\hat{H}_\text{trial}=\frac{i\hbar}{T}\log(\hat{\mathcal{F}}_\text{trial})$. Thus, different equivalent effective Hamiltonians correspond to the same effective unitary dynamics. To find the desired effective Hamiltonian, we optimize the real part of the trace
%%%
\begin{equation}
F=\frac{1}{D_{M,L}}\Re\left[\text{tr}(\hat{\mathcal{F}}_{\text{target}}^\dag \hat{\mathcal{F}}_{\text{trial}})\right] \, .
\end{equation}
%%%
To  numerically solve the control problem, the time evolution is discretized into $N$ steps of length $\tau$. The propagator of the j-th time step $t_j=j\tau$ is given by
\begin{equation}
\hat{U}_j=e^{-i \hat{H}_u(t_j)\tau/\hbar} \, ,
\end{equation}
with the total Hamiltonian for time step $t_j$ as $\hat{H}_u(t_j)=\hat{H}_\text{d}+\sum_{k=1}^R u_k(t_j)\hat{V}_k$. The total time evolution is then given by $\hat{U}(T;0)=\hat{U}_N\hat{U}_{N-1}\cdots\hat{U}_{2}\hat{U}_{1}$.
The control parameters $u_k(t_j)$ are initially chosen at random, and are to be optimized to achieve the desired target unitary $\hat{\mathcal{F}}_{\text{target}}$. The method of choice here is gradient ascent. The idea is to optimize the fidelity by taking small steps in the control parameters by following the gradient of the fidelity.
The gradients $\frac{\partial}{\partial u_k(t_j)} F(\hat{U}_j)$ can be calculated by invoking the spectral theorem \cite{machnes2011comparing}. Then, all the control parameters are updated at the same time. This procedure is repeated until convergence. We are using the L-BFGS-B algorithm within the Qutip implementation \cite{johansson2013qutip}. As gradient descent can become stuck in local minima, we repeat the optimization several times with different random initial conditions. 

%%%
\begin{figure*}
 \centering
 \includegraphics[width=0.85\textwidth]{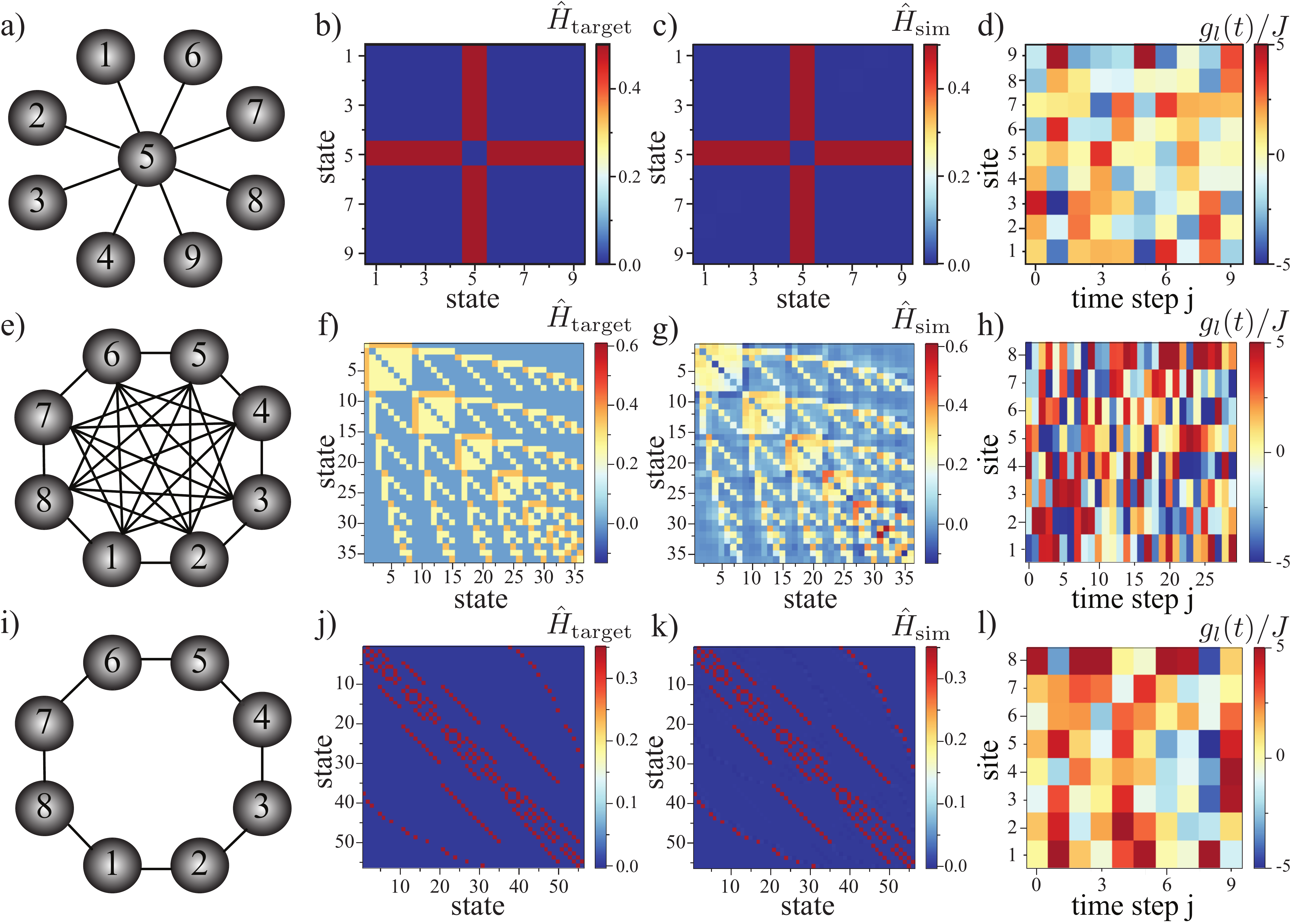}
	\caption{Hamiltonian engineering for $M$ excitations in $L$ sites. Panels a)-d) depict the case of a single excitation $M=1$ in $L=9$ sites, where we simulate a network with star topology. Panels  e)-h) illustrate the case of $M=2$ in $L=8$ sites, where we engineer a fully connected graph. Correspondingly, panels i)-l) show the case of $M=3$ excitations in $L=8$ sites that is used to engineer a ring.
a), e) and i) Show the graph representing the target topology that we aim to achieve, starting from a linear spin chain with nearest-neighbohrs coupling. b), f) and j) Show a matrix plot of the target Hamiltonian corresponding to a), e) and i). c), g) and k) Show the matrix representation the effective Hamiltonian obtained after optimization of the driving sequence. Correspondingly, panels  d,h,l) illustrate the driving sequence of the onsite energies $-5J<g_l(t)<5J$ for each time step $j$ with driving periods $T=10/J$, $T=15/J$ and $T=10/J$, respectively.
 }
\label{Fig2}
\end{figure*}
 %%%	

%%%%%
\subsection{The family of effective Hamiltonians that can be simulated with our scheme}
%%%%%
As a result of the optimization, we obtain a driving sequence that leads to an optimal unitary $\hat{\mathcal{F}}_{\text{sim}}=\hat{U}_{\text{sim}}(T)=\exp(-\mathrm{i}\hat{H}_{\text{sim}}T/\hbar)$. The effective Hamiltonian $\hat{H}_{\text{sim}}$ is very close to the target Hamiltonian $\hat{H}_{\text{target}}$ and can be used to simulate it.  An alternative way  to find the driving sequence is to use machine learning methods, which have been successfully applied  to quantum dynamics~\cite{haug2020classifying} and quantum control~\cite{Bukov2018,haug2019engineering,Niu2019}.

At each time $t$, the Hamiltonians~\eqref{eq:HamiltonianResonatorArray}~and~\eqref{eq:HamiltonianHarcoreBoson} are strictly limited to the linear chain connectivity, which is usually fixed at the fabrication of the device. However, after the optimization described above we can simulate a target Hamiltonian
 %%%
\begin{align}
\label{eq:EffHamiltonianBoson}
\hat{H}_{\text{target}}&=\hbar\sum^{L}_{l=1}\left[G_{l} \hat{n}_{l}+\frac{U}{2} \hat{n}_{l}(\hat{n}_{l}-1)\right]+\hbar \sum^{L}_{l,m=1}K_{l,m}(\hat{a}^{\dagger}_{l}\hat{a}_{m}+\text{h.c})
\ 
 \end{align}
%%%
with an arbitrary connectivity.
The parameters $G_{l}$ are the effective angular frequencies of the qubits and $K_{l,m}$ contains information about the engineered connectivity of the sites, as depicted in Fig.\ref{Fig1}c).  There one can see that by driving a simple one-dimensional chain with nearest-neighbor couplings $J_l(t)$, one can engineer arbitrary couplings $K_{l,m}$ between arbitrary sites $l$ and $m$. Further, due to the nature of the hardcore bosons, it turns out that we can achieve spin Hamiltonians with arbitary connectivity
 %%%
\begin{align}
\label{eq:EffHamiltonianHarcoreBoson}
\hat{H}_{\text{target}}&=\frac{\hbar}{2}\sum^{L}_{l=1}G_{l} \sigma^z_{l}+\frac{\hbar}{2} \sum^{L}_{l,m=1}K_{l,m}(\sigma^x_{l}\hat{O}_{l,m}\sigma^x_{m}+\sigma^y_{l}\hat{O}_{l,m}\sigma^y_{m})
 \ ,
 \end{align}
%%%
where $\hat{O}_{l,m}=\sigma^z_{l+1}\sigma^z_{l+2}\cdots\sigma^z_{m-2}\sigma^z_{m-1}$ is a nonlocal operator keeping track of the Jordan-Wigner strings (see methods)~\cite{nagaosa1999quantum}.

%%%%%
\section{Simulating arbitrary connectivities by driving a one-dimensional lattice with $M$ excitations in $L$ sites}
\label{SecIV}
%%%%%
Let us now present some specific examples that illustrate how we can simulate a nontrivial topology starting from a simple one. Let us start with the simplest example of a single excitation $M=1$ in an array of $L=9$ sites, where $D_{1,9}=9$ is the dimension of the Hilbert space. In the single-excitation subspace,
we consider the basis $\ket{1_l}=\ket{0,0,\ldots,1_l,0,\dots,0}$ with $l=1,\dots,L$, where $\ket{1_l}$ denotes an excitation at the $l$-th site.
In this case, we use Hamiltonian~\eqref{eq:HamiltonianResonatorArray}, but as we are working with a single excitation, it is also possible to use Hamiltonian \eqref{eq:HamiltonianHarcoreBoson} to obtain the same results. Consider that we want to simulate a star network topology of the effective Hamiltonian~\eqref{eq:EffHamiltonianBoson} or \eqref{eq:EffHamiltonianHarcoreBoson} as depicted in Fig.~\ref{Fig2}~a). In Fig.~\ref{Fig2}~b) we illustrate the matrix representation of target Hamiltonian in the single-excitation basis $\ket{1_l}$. As we discussed before, it is non-trivial to obtain the right sequence of driving parameters $\bar{g}^j_l$ and $\bar{J}^j_l$ for sites $l$ and time $t_j$ in order to achieve the target Hamiltonian. Using GRAPE, we obtain an effective Hamiltonian as in Fig.~\ref{Fig2}~c) that is very close to the target. This can be done experimentally by driving the onsite energies $\bar{g}^j_l$ of the chain, while keeping the couplings constant $(\bar{J}^j_l=J)$. The driving protocol is shown in Fig.~\ref{Fig2}~d).

%%%
\begin{figure*}
 \centering
 \includegraphics[width=0.90\textwidth]{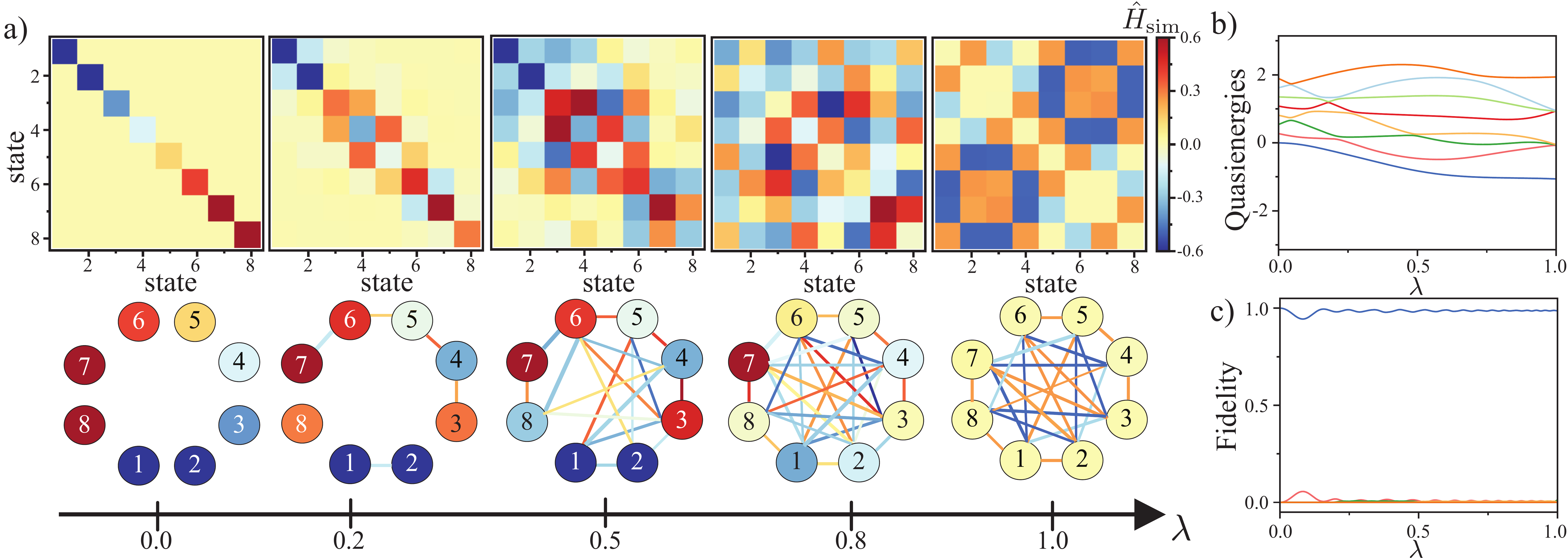}
	\caption{Hamiltonian engineering for 3-SAT problems with three-body interactions and adiabatic deformation between effective Hamiltonians.  a) Shows the adiabatic process $\hat{H}_{\text{adiabatic}}(\lambda)=(1-\lambda)\hat{H}_{\text{diag}}+\lambda \hat{H}_{\text{SAT}}$ that allows us to deform a diagonal Hamiltonian into a 3-SAT Hamiltonian $\hat{H}_{\text{SAT}}=4\hbar(\mu^x_{1}-\mu^x_{2}+\mu^x_{3}-2\mu^x_{1}\mu^x_{2}+2\mu^x_{2}\mu^x_{3}+\mu^x_{1}\mu^x_{2}\mu^x_{3})$ as a function of the parameter $\lambda$. The upper and lower panels depict the effective Hamiltonians and the graph representations, respectively. b) Shows the quasienergies as a function of $\lambda$. c) Illustrates the fidelity between the state of system and instantaneous eigenstates of effective Hamiltonian. We can clearly see that under an adiabatic deformation in the space of effective Hamiltonians, the system remains in its ground state, as expected. The adiabatic evolution happens over a total time of $T_\text{total}=200 T=1276/J$, where $T=6.38/J$ is the period of the drive. In  appendix~\ref{AppendixD}, we provide details on the driving sequence that is required to simulate the 3-SAT Hamiltonian $\hat{H}_{\text{SAT}}$. } 
\label{Fig3}
\end{figure*}
 %%%

Next let us move beyond the single excitation subspace and effectively expand the computational space by considering $M=2$ excitations in $L=8$ sites. In this subspace, we consider the basis states $\ket{1_l,1_m}=\ket{0,0,\ldots,1_l,0,\dots,1_m,0,\ldots,0}$ with $l,m=1,\dots,L$. Here $\ket{1_l,1_m}$ denotes the state of two excitations at sites $l$ and $m$. Note that here we consider $U\sim g_{l}, J_{l}$, which is far from the hardcore boson regime and thus, we allow for states $\ket{2_l}=\ket{0,0,\ldots,2_l,0,\ldots,0}$ describing two excitations being at the same site. In this case, the dimension of the Hilbert space is $D_{2,8}=36$. Our aim here is to simulate a target Hamiltonian~\eqref{eq:HamiltonianResonatorArray} describing a device with all-to-all connectivity, as it is depicted in Fig.~\ref{Fig2}~e). Figs.~\ref{Fig2}~f)~and~g) show the target and effective Hamiltonians, respectively. The driving sequence of the onsite energies that is sufficient to simulate this topology is illustrated in Figs.~\ref{Fig2}~h). In contrast to the single particle case, here we drive both the onsite energies $\bar{g}^j_l$ as well as the  couplings $\bar{J}^j_l$. In  appendix~\ref{AppendixC} we show the driving sequence of the couplings $\bar{J}^j_l$.

In the hardcore boson regime we can simulate a target Hamiltonian~\eqref{eq:EffHamiltonianHarcoreBoson} describing $M=3$ excitations in a ring with $L=8$ sites [see Fig.~\ref{Fig2}~i)]. In this regime, however, there is a restriction on the type of Hamiltonians that can be simulated (see methods), because the doubly and triply occupied states are forbidden due to the large nonlinearity. For this reason, we work with the basis $\ket{1_k,1_l,1_m}=\ket{0,\ldots,1_k,0,\dots,1_l,0\dots,1_m,\ldots,0}$ with $k\neq l\neq m$ and the dimension of the Hilbert space is $D_{3,8}=55$. Figs.~\ref{Fig2}~j)~and~k) show the target and effective Hamiltonians, respectively and the driving sequence is illustrated in Fig.~\ref{Fig2}~l). In  appendix~\ref{AppendixE} we show the scaling of our method in the hardcore boson regime as we increase the number of particles~$M$.

%%%%%
\section{Hamiltonian engineering and adiabatic deformation for 3-SAT problems}
\label{SecV}
%%%%%
We have shown that by exploiting tools of optimization, we can simulate complex topologies at stroboscopic times by using Hamiltonians~\eqref{eq:HamiltonianResonatorArray}~and~\eqref{eq:HamiltonianHarcoreBoson}. Next, let us explore a practical application of our approach as a stroboscopic simulator of a SAT solver for combinatorial optimization. The term SAT itself refers to satisfiability of equations involving boolean variables and some variants of SAT problems are hard~\cite{battaglia2005optimization}. 3-SAT problems are of utmost importance because any $k$-SAT problem can be decomposed into a sequence of $3$-SAT instances. The simplest $3$-SAT problems involve $3$-body interactions~\cite{battaglia2005optimization}. To generate this kind of interactions in near-term  analogue quantum devices can be quite challenging, due to the restrictions of the hardware. Therefore, we expect our method to provide a route to design the desired target Hamiltonian. Next,  to demonstrate the versatility of our approach, we explore an example of a $3$-SAT problem to solve the clauses $1+a_{2}+a_{3}+a_{1}a_{3}=0$, $1+a_{1}+a_{3}+a_{1}a_{2}=0$, and $a_{1}+a_{2}+a_{2}a_{3}=0$, where $a_1, a_2$ and $a_3$ are boolean variables. The latter is a typical
example of a system of equations that arises when performing cryptanalysis
of block ciphers or hash functions~\cite{GraPanThoUniPer2019}. By using algebraic tools~\cite{MulPanHanFin2013}, it is possible to map this combinatorial problem to a spin Hamiltonian 
 %%%
\begin{align}
\label{eq:3SATHamiltonian}
\hat{H}_{\text{SAT}}&=\hbar\omega(\mu^x_{1}-\mu^x_{2}+\mu^x_{3}-2\mu^x_{1}\mu^x_{2}-2\mu^x_{1}\mu^x_{3}+\mu^x_{1}\mu^x_{2}\mu^x_{3})
 \end{align}
%%%
with three-body interactions, where $\mu^{\alpha}_l$ are Pauli matrices  and $\omega$ is a parameter with units of angular frequency ( see appendix~\ref{AppendixD}).
Thus, by finding its ground state, we can obtain the solution of the optimization problem, which in this case, corresponds to the triplet $(a_1, a_2, a_3)$ satisfying  the 3 clauses discussed above. In the eigenbasis of $\mu^z_{l}$, the matrix representation of Hamiltonian~\eqref{eq:3SATHamiltonian} is a $8\times8$ matrix, as it is depicted in Fig.\ref{Fig3}~a). It can also be represented as a graph with high-connectivity, which is difficult to implement experimentally. With our approach, we obtain a driving sequence that stroboscopically simulate the target Hamiltonian $\hat{H}_{\text{SAT}}$, as we discuss in appendix~\ref{AppendixD}. This can be mapped to a single-excitation in terms of Hamiltonian~\eqref{eq:EffHamiltonianBoson} and~\eqref{eq:EffHamiltonianHarcoreBoson}.

\begin{figure*}
	\centering
	 \includegraphics[width=0.90\textwidth]{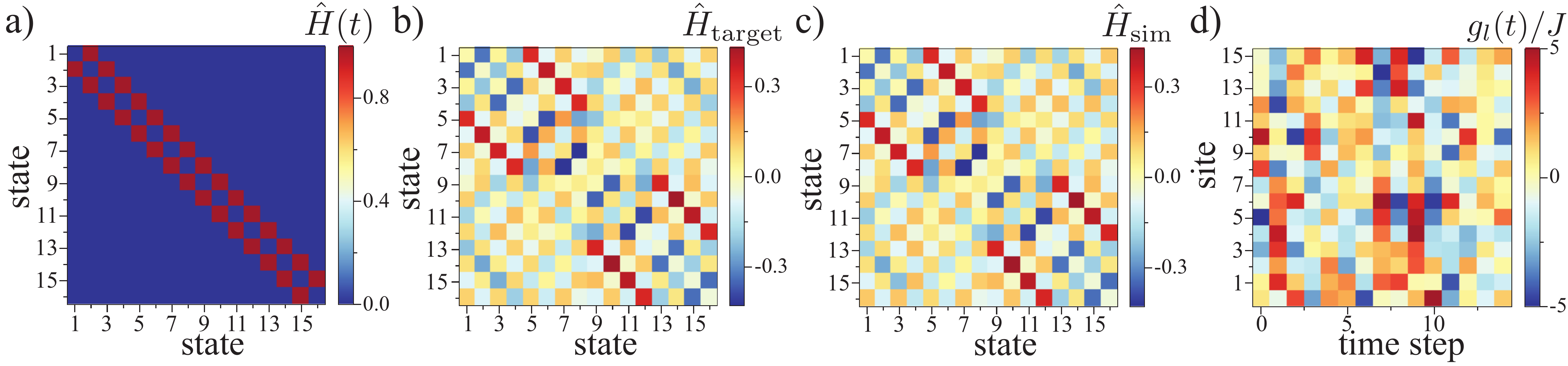}
	\caption{Simulation of LiH with Floquet engineering. The Molecule can be mapped to a Hilbert space with dimension $D_{1,16}$. a) Depicts the instantaneous nearest-neighbor Hamiltonian of a linear chain with $L=16$ sites and one excitation ($M=1$). b) Shows the target Hamiltonian, while c) depicts the effective Hamiltonian generated by driving local potential of linear chain. d) Illustrates the driving sequence $-5J<g_l<5J$ applied on the diagonal terms between  to generate effective Hamiltonian.
	}
	\label{Fig4}
\end{figure*}

Next, we explore an adiabatic deformation in the space of effective Hamiltonians by slowly switching on the driving protocol~\cite{Tanaka2010}. In this way, we can deform a trivial diagonal Hamiltonian $\hat{H}_{\text{diag}}$ into the Hamiltonian~\eqref{eq:3SATHamiltonian} by using the Hamiltonian $\hat{H}_{\text{adiabatic}}(\lambda)=(1-\lambda)\hat{H}_{\text{diag}}+\lambda \hat{H}_{\text{SAT}}$, where $0<\lambda<1$ is slowly modulated in time.  Fig.\ref{Fig3}~a) depicts the adiabatic deformation of the Hamiltonian $\hat{H}_{\text{adiabatic}}(\lambda)$ and its corresponding representation in terms of graphs for different values of $\lambda$. During the adiabatic deformation, the quasienergies exhibit multiple anticrossings, as depicted in Fig.\ref{Fig3}~b), but the system remains in its lowest quasienergy state, as illustrated in Fig.\ref{Fig3}~c). 
By finding the ground state when $\lambda=1$, we obtain a solution $(a_1, a_2, a_3)=(1,0,0)$ satisfying the three clauses discussed above. For more details, we refer the reader to the appendices. In the next section, we provide an example of how to use our approach in quantum Chemistry to simulate a LiH molecule at bond distance, where nontrivial many-body terms can directly encoded into the effective Hamiltonian.

\section{Hamiltonian engineering in quantum chemistry: Stroboscopic simulation of LiH}
\label{SecVI}

In the previous section we discussed an application of our method to simulate optimization problems. Here, we explore yet another application to stroboscopically simulate Hamiltonians in quantum chemistry. This is one of the most relevant near term applications of the existing NISQ devices.

Understanding biological processes and designing new pharmaceutical products is of utmost importance in todays age. To accomplish task, we have to understand the inner workings of atoms, molecules and proteins better by calculating their properties such as their quantum mechanical configuration and dynamics from first principles.  However, the simulation of molecules is a difficult task for classical computers as the computational resources scale exponentially with the number of electrons that have to be calculated. Quantum computers do not suffer from this scaling and thus promise to be able to calculate large molecules which are intractable with classical computation. As benchmark, small molecules can be calculated already with state-of-the-art quantum computers \cite{kandala2017hardware}. The Hamiltonian that describes the molecule includes non-local interactions between multiple qubits that should be controlled. Most quantum computers support only two-body interactions and thus require multiple operations just to fulfill a single non-local operation acting on multiple qubits. 

With Floquet engineering, all those interactions including the complex n-body qubit interactions can directly encoded into the effective Hamiltonian.
We demonstrate this by simulating a LiH molecule at bond distance in Fig.\ref{Fig4}. The corresponding parameters of the Hamiltonian can be found in \cite{kandala2017hardware}. They have been obtained by the STO-3G basis. It approximates the atomic orbitals with three Gaussians to obtain  the one and two-electron integrals for the electron interactions. For the LiH molecule, the 1s orbital of H and the 1s, 2s and 2p$_x$ and 2p$_z$ were assumed to be occupied, all other orbitals are assumed to be empty. By including the parity symmetries, the LiH molecule can be described effectively by 4 qubits or a Hilbert space of dimension $D_{1,16}=16$. It consists of 99 coupling terms expressed as a product of different Pauli operators (see appendix~\ref{AppendixF}). 
Fig.\ref{Fig4}~a) illustrate the instantaneous Hamiltonian of the device. We use a one-dimensional chain with a single excitation $M=1$ in $L=16$ sites. Figs.\ref{Fig4}~b)~and~c) show the good agreement between the target and simulated Hamiltonian, respectively. The stroboscopic simulation of the LiH molecule can be efficiently achieved by driving the onsite energies, as depicted in Fig.\ref{Fig4}~d).

%%%%%
\section{Conclusions}
\label{SecVII}
%%%%%

In this work we have demonstrated that one can simulate a target Hamiltonian with arbitrary connectivity by driving a lattice with a simple fixed topology. To achieve this, we exploited tools of optimization to engineer a periodic drive that allows us to simulate the target Hamiltonian at stroboscopic times. We demonstrate the versatility of our approach by simulating star, all-to-all and ring connectivities. Further, we propose a practical application of our methods: We simulated the 3-SAT Hamiltonian, a paradigmatic model of combinatorial optimization. In terms of Pauli matrices, the 3-SAT model involves 3-body interactions, which are difficult to realize experimentally. We envision potential applications of our results to quantum chemistry and quantum simulation in noisy intermediate-scale quantum (NISQ) devices.

%%%%%
\section{Acknowledgements}
%%%%%
We thank M. P. Estarellas, A. Sakurai, L. Wright for valuable discussions and Y. Sato for proof-reading the manuscript.
This work was supported in part by the Japanese MEXT Quantum Leap Flagship Program (MEXT Q-LEAP), Grant Number JP-MXS0118069605, the MEXT KAKENHI Grant-in-Aid for Scientific Research on Innovative Areas Science of hybrid quantum systems Grant No.15H05870 and the JSPS KAKENHI Grant No. 19H00662. The computational work for this article was partially per-formed on resources of the National Supercomputing Centre, Singapore (https://www.nscc.sg).

\appendix

\section{Floquet theory and stroboscopic dynamics}
\label{AppendixA}
In our manuscript we focus on time periodic Hamiltonians $\hat{H}(t+T)=\hat{H}(t)$ where $T$ is the drive period. Due to the periodicity of the Hamiltonian, the most relevant information is contained in the Floquet operator $\hat{\mathcal{F}}=\hat{U}(T;0)=\hat{\mathcal{T}}\exp{\left[-\mathrm{i}/\hbar\int_0^T\hat{H}(s) d s\right]}$, which is the evolution operator  within one period of the drive. In the previous equation, we need to use the time-ordering operator $\hat{\mathcal{T}}$. By solving the eigenvalue problem $\hat{\mathcal{F}}|\Phi_{\alpha}\rangle=e^{-\mathrm{i}\varepsilon_{\alpha}T/\hbar}\Phi_{\alpha}\rangle$, one can obtain the most relevant information for the dynamics. The eigenvectors $|\Phi_{\alpha}\rangle$ are known as the Floquet states and $-\hbar\pi/T\leq\varepsilon_{\alpha}\leq\hbar\pi/T$ are the quasienergies. As the Floquet operator is unitary, it is possible to define an effective Hamiltonian $\hat{H}_{\text{target}}$ such that  $\hat{\mathcal{F}}=\exp(-\mathrm{i}\hat{H}_{\text{target}}T/\hbar)$. At stroboscopic times $t_n=nT$ the effective Hamiltonian is the generator of the dynamics. To be more concrete, given an initial state $\ket{\psi(0)}$, its time evolution at times $t_n=nT$ is given by $\ket{\psi(n)}=\hat{\mathcal{F}}^n\ket{\psi(0)}=\exp(-\mathrm{i}\hat{H}_{\text{target}}nT/\hbar)\ket{\psi(0)}$, which looks exactly as the evolution under a time independent Hamiltonian.

\section{Hardcore bosons and the Jordan-Wigner transformation}
\label{AppendixB}
In this section, we discuss in detail the family of target Hamiltonians that can be simulated using the Hamiltonian
%%%
\begin{align}
\label{eq:SpinHamiltonian}
\hat{H}(t)&=\frac{\hbar}{2}\sum^{L}_{l=1}g_{l}(t) \sigma^z_{l}+\frac{\hbar}{2} \sum^{L-1}_{l=1}J_l(t)(\sigma^x_{l}\sigma^x_{l+1}+\sigma^y_{l}\sigma^y_{l+1})\ 
\end{align}
%%%
by describing microwave photons in the hardcore boson regime $U\gg g_l,J_l$. In this case, one can use the Jordan-Wigner transformation~\cite{nagaosa1999quantum}

%%%%
\begin{equation}
 \sigma^z_l = -f^{\dagger}_l e^{\mathrm{i}\hat{\Phi}_l}-f_l e^{-\mathrm{i}\hat{\Phi}_l},\,\  \sigma^y_l = -\mathrm{i}f^{\dagger}_l e^{\mathrm{i}\hat{\Phi}_l}+\mathrm{i}f_l e^{-\mathrm{i}\hat{\Phi}_l}, \,\   \sigma^x_l =2f^{\dagger}_l f_l-1 \ ,
\end{equation}
%%%
with $\hat{\Phi}_l=\sum_{j < l}f^{\dagger}_j f_j$ to map the spin Hamiltonian to the fermionic representation
%%%
\begin{equation}
\label{eq:FermionicHamiltonian}
\hat{H}(t)=\hbar\sum^{L}_{l=1}\left[g_{l}(t)f^{\dagger}_{l}f_{l}+J_l(t)(f^{\dagger}_{l}f_{l+1}+\text{h.c.})\right] \, ,
\end{equation}
%%%
where $f^{\dagger}_{j}$ ($f_{j}$) are fermionic creation (annihilation) operators. The fermionic representaion is versatile, because it gives us a canonical form of the target Hamiltonians that can be achieved by applying a periodic drive
%%%
\begin{equation}
\label{eq:EffectiveFermionicHamiltonian}
\hat{H}_{\text{target}}=\hbar\sum^{L}_{l=1}\left[G_{l}f^{\dagger}_{l}f_{l}+K_{l,m}(f^{\dagger}_{l}f_{m}+\text{h.c.})\right] \ .
\end{equation}
%%%
Crucially, the effective Hamiltonian allows for long-range hopping of Jordan-Wigner fermions. Due to the nonlocal character of the Jordan Wigner transformation, these long-range hopping become highly nonlocal terms in the spin representation, as follows
 %%%
\begin{align}
\label{eq:MethodsEffHamiltonianHarcoreBoson}
\hat{H}_{\text{target}}&=\frac{\hbar}{2}\sum^{L}_{l=1}G_{l} \sigma^z_{l}+\frac{\hbar}{2} \sum^{L}_{l,m=1}K_{l,m}(\sigma^x_{l}\hat{O}_{l,m}\sigma^x_{m}+\sigma^y_{l}\hat{O}_{l,m}\sigma^y_{m})
 \ ,
 \end{align}
%%%
where $\hat{O}_{l,m}=\sigma^z_{l+1}\sigma^z_{l+2}\cdots\sigma^z_{m-2}\sigma^z_{m-1}$. To simulate a ring, as we discussed in the main text, we require effective couplings between nearest neighbohrs $K_{l,l+1}$  and to have a coupling $K_{1,L}$. In terms of the spin representation, this leads to an effective Hamiltonian
%%%
\begin{align}
\label{eq:RingHarcoreBoson}
\hat{H}_{\text{target}}&=\frac{\hbar}{2}\sum^{L}_{l=1}G_{l} \sigma^z_{l}+\frac{\hbar}{2} \sum^{L-1}_{l=1}K_{l,l+1}(\sigma^x_{l}\sigma^x_{l+1}+\sigma^y_{l}\sigma^y_{l+1})
\nonumber\\
&+K_{1,L}(\sigma^x_{1}\hat{O}_{1,L}\sigma^x_{L}+\sigma^y_{1}\hat{O}_{1,L}\sigma^y_{L})
 \ .
 \end{align}
%%%

%%%%
\section{Instantaneous Hamiltonians and driving protocols for star, all-to-all and ring connectivities}
\label{AppendixC}
%%%%
%%%%%%%%
\begin{figure}
	\centering
	 \includegraphics[width=0.50\textwidth]{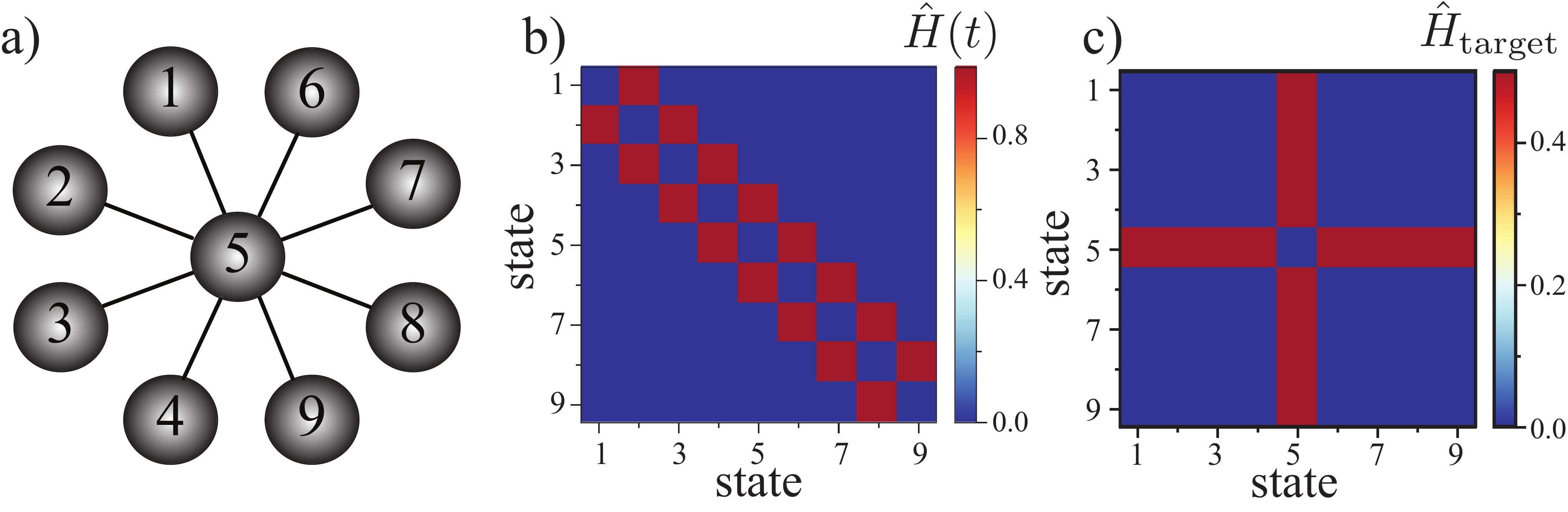}
	\caption{Star connectivity in the single-excitation subspace:Instantaneous and Target Hamiltonian.
	a) Illustrates the target connectivity that we want to simulate with a single excitation $M=1$ in $L=9$ sites. b) Depicts the matrix representation of the Hamiltonian of the one-dimensional quantum processor at each time $t$ in the basis $\ket{1_l}=\ket{0,0,\ldots,1_l,0,\dots,0}$. From this one can clearly see that each site is coupled to nearest neighbors in a chain with open boundary conditions. c) Depicts the target effective Hamiltonian, where site $l=5$ is connected to all the other sites, thus simulating a star connectivity of the device. The period of the driving sequence $-5J<g_l<5J$ is set to be $T=10/J$ and it is divided into $N=10$ time steps. It is important to remark that in this case we drive only the onsite energies and we keep the couplings $J_l=J$ between the sites constant.
	}
	\label{Fig1S}
\end{figure}
%%%%%%%%
Our aim in this section is to provide additional details on the simulated connectivities that we describe in the main text. At each time, the device has a connectivity of a chain with open boundary conditions. Our objective is to apply a periodic drive to the quantum processor to simulate an effective Hamiltonian at stroboscopic times. By means of the periodic driving sequence, one can simulate couplings that are absent in the original quantum processor. 
By using superconducting qubits, the driving sequence can be achieved by applying pulses through the Z lines of a superconducting processor, which can be done by rapidly bringing the qubit from its iddle frequency to the working point. In this way, one can generate a local modulation $-5J<g_l<5J$ of the onsite energies. In superconducting processors such as gmons, the couplings are tunable and the couplings can me modulated between positive and negative values  with $-J<J_l<J$. 

%%%%%%%%
\begin{figure}
	\centering
	\includegraphics[width=0.50\textwidth]{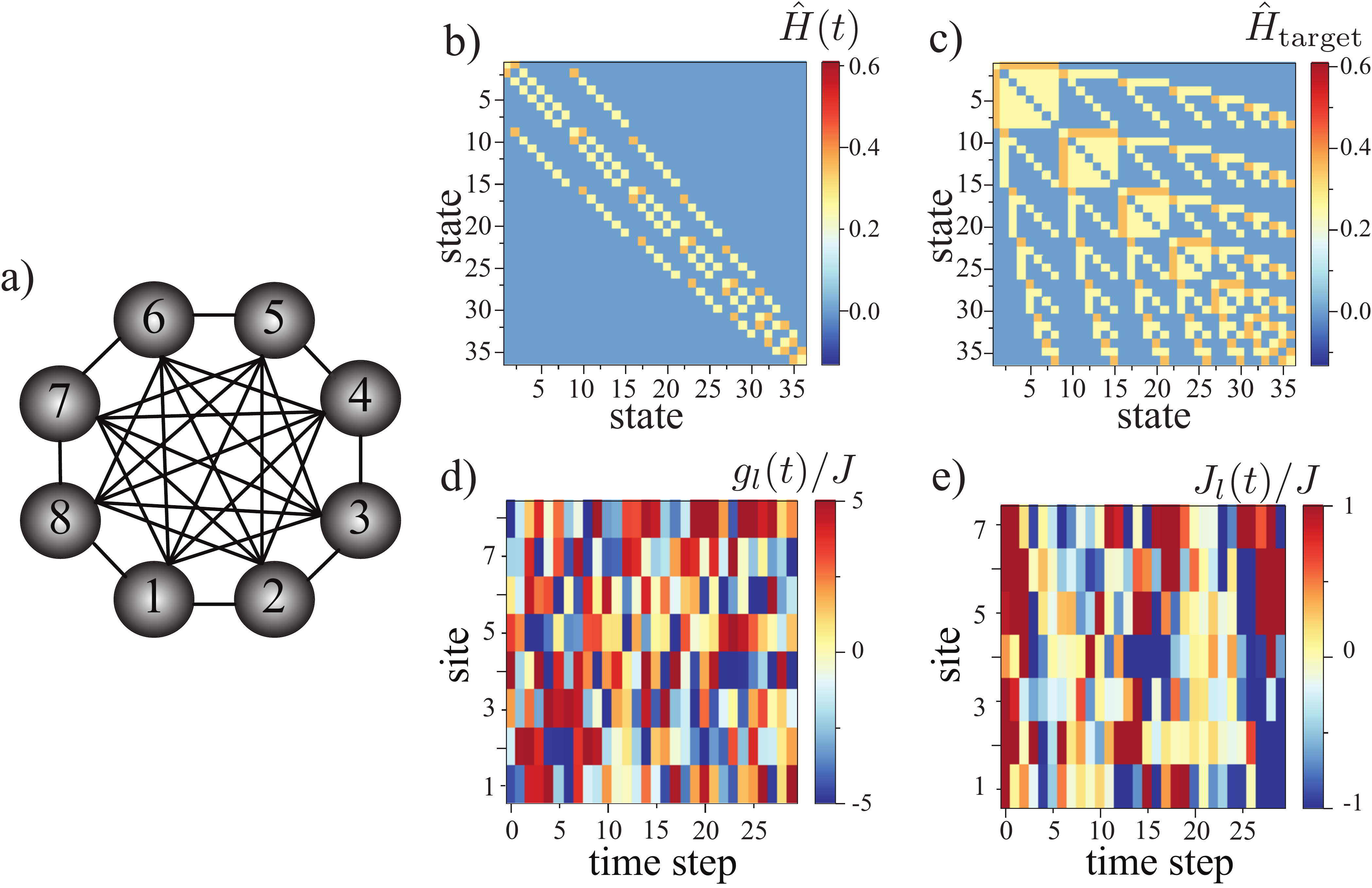}
	\caption{All-to-all connectivity: Instantaneous and Target Hamiltonian. a) Illustrates the target connectivity that we want to simulate with two excitations $M=2$ in $L=8$ sites. b) Depicts the matrix representation of the Hamiltonian of the one-dimensional quantum processor at each time $t$ in the basis $\ket{1_l,1_m}=\ket{0,0,\ldots,1_l,0,\dots,1_m,0,\ldots,0}$. c) Depicts the target effective Hamiltonian. In d) and e) we also depict the driving protocol on the onsite energies $g_l(t)$ and the couplings $J_l(t)$ obtained using GRAPE. The period of the driving sequence is set to be $T=15/J$ and the driving sequence $-5J<g_l<5J$, $-J<J_l<J$ is divided into $N=30$ time steps. We set the nonlinearity to be $U=4J$.
	}
	\label{Fig2S}
\end{figure}
%%%%%%%%

Let us start by discussing the case of a single excitation  $M=1$ in an array of $L=9$ sites. In this case, the dimension of the Hilbert space is $D_{1,9}=9$ and we consider the basis $\ket{1_l}=\ket{0,0,\ldots,1_l,0,\dots,0}$ with $l=1,\dots,L$, where $\ket{1_l}$ denotes an excitation at the $l$-th site. In the single-excitation subspace, the statistics of the excitations do not play any role. 
%%%%%%%%
\begin{figure}
	\centering
	\includegraphics[width=0.50\textwidth]{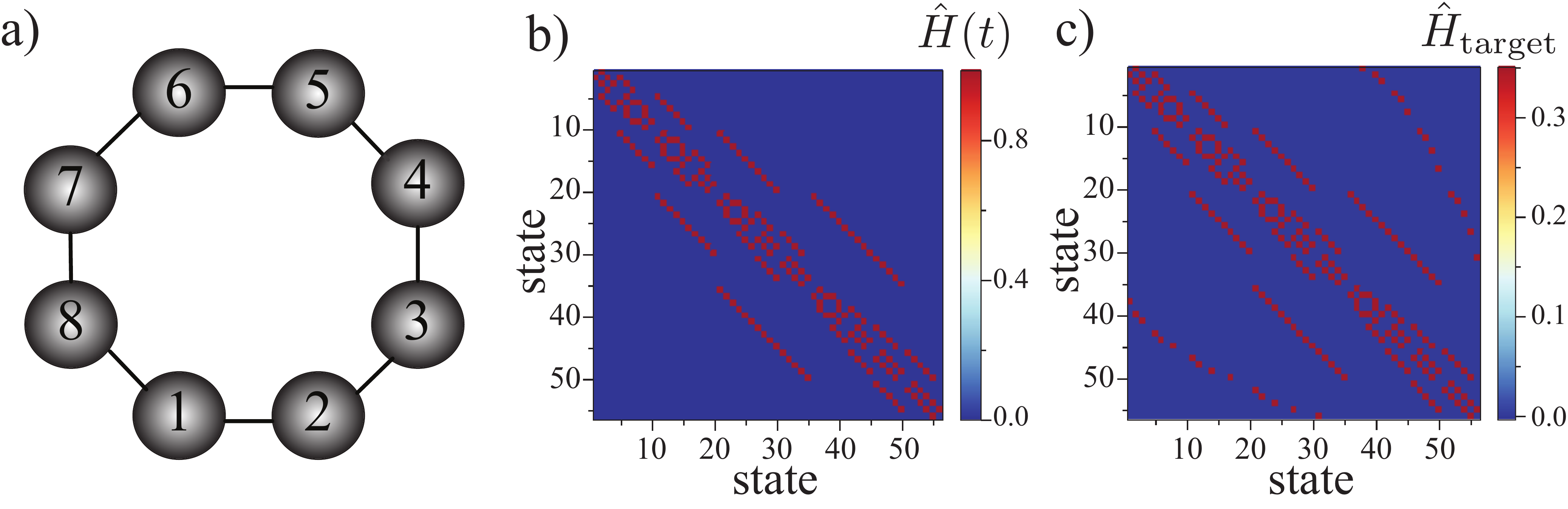}
	\caption{Ring connectivity: Instantaneous and Target Hamiltonian. a) Illustrates the target connectivity that we want to simulate with two excitations $M=3$ in $L=8$ sites. b) Depicts the matrix representation of the Hamiltonian of the one-dimensional quantum processor at each time $t$ in the basis  $\ket{1_k,1_l,1_m}=\ket{0,\ldots,1_k,0,\dots,1_l,0\dots,1_m,\ldots,0}$ with $k\neq l\neq m$ . c) Depicts the target effective Hamiltonian. The period of the driving sequence is set to be $T=10/J$ and the driving sequence $-5J<g_l<5J$ is divided into $N=10$ time steps. To simulate the ring connectivity, we drive only the onsite energies and we keep the couplings $J_l=J$ between the sites constant.}
	\label{Fig3S}
\end{figure}
%%%%%%%%
Therefore, we can either work with the Hamiltonian  Eq.$(1)$ for bosons  or Hamiltonian Eq.$(2)$ for hardcore bosons in the main text to simulate the star connectivity depicted in Fig.~\ref{Fig1S}~a). The matrix representation of the instantaneous Hamiltonian $\hat{H}(t)$ is illustrated in Fig.~\ref{Fig1S}~b). As we describe in Fig.2 d) of the main text, by applying a sequence of pulses to the sites $l$ of the chain, one can simulate the target Hamiltonian in Fig.~\ref{Fig1S}~c). To simulate the star connectivity, it is enough to locally drive the onsite energies with a modulation strength $-5J<g_l<5J$.

Next, let us discuss more details on the case of two excitations $M=2$ in $L=8$ sites. Our goal is to provide more details on the results presented in Figs. 2 e)-h) in the main manuscript. As now the statistics of the excitations plays an important role, we restrict ourselves to the Hamiltonian Eq.(1) of the main text for interacting bosons. In the subspace with two excitations, we consider the basis states $\ket{1_l,1_m}=\ket{0,0,\ldots,1_l,0,\dots,1_m,0,\ldots,0}$ with $l,m=1,\dots,L$. Here $\ket{1_l,1_m}$ denotes the state of two excitations at sites $l$ and $m$.
We set the interaction strength $U=4J$ and as we are far from the hardcore boson regime, we allow for states $\ket{2_l}=\ket{0,0,\ldots,2_l,0,\ldots,0}$ describing two excitations being at the same site. In this case, the dimension of the Hilbert space is $D_{2,8}=36$. Our objective is to simulate all-to-all connectivity of the quantum processor as illustrated in Fig.~\ref{Fig2S}~a). At each time, the matrix representation of the Hamiltonian in the subspace with two excitations is shown in Fig.~\ref{Fig2S}~b). Correspondingly, Fig.~\ref{Fig2S}~c) depicts the matrix representation of the target Hamiltonian. To be able to simulate the target Hamiltonian, we have to apply a periodic driving sequence  $-5J<g_l<5J$, $-J<J_l<J$ with a period $T=15/J$ and within a period, the sequence is divided into $N=30$ time steps (or pulses). As we discussed at the beginning of this section. This type of driving sequences is within reach with available technologies in superconducting qubits arrays. 

%%%%%%%%
\begin{figure}
	\centering
	\includegraphics[width=0.5\textwidth]{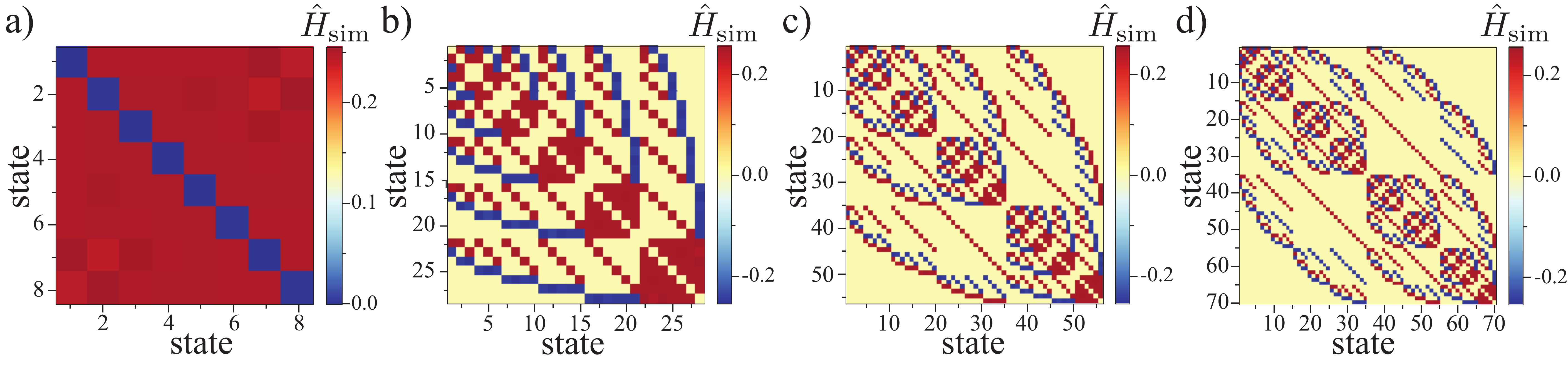}
	\caption{All-to-all connectivity using multiple excitations in the hardcore boson regime.
	a)-d) Effective Hamiltonian for all-to-all coupling for varying excitation number $M$. Hamiltonian corresponds to all-to-all coupled non-interacting fermion Hamiltonian. a) $M=1$ in the basis $\ket{1_l}$, b) $M=2$ in the basis $\ket{1_l,1_m}$, c) $M=3$ in the basis $\ket{1_k,1_l,1_m}$, and d) $M=4$ in the basis $\ket{1_k,1_l,1_m,1_o}$ such that $k\neq l\neq m\neq 0$. The period of the driving sequence is set to be $T=10/J$ and the driving sequence $-5J<g_l<5J$ is divided into $N=10$ time steps. To simulate the ring connectivity, we drive only the onsite energies and we keep the couplings $J_l=J$ between the sites constant.
	}
	\label{Fig4S}
\end{figure}
%%%%%%%%

Now we provide more information of the results obtained in the hardcore boson regime $U\gg J_l, g_l$. In this regime, we simulate a ring connectivity by using   $M=3$ excitations in $L=8$ sites. Due to the large nonlinearity, the doubly and triply occupied states are forbidden and for this reason, we work with the basis $\ket{1_k,1_l,1_m}=\ket{0,\ldots,1_k,0,\dots,1_l,0\dots,1_m,\ldots,0}$ with $k\neq l\neq m$ and the dimension of the Hilbert space is $D_{3,8}=55$. In Figs.~\ref{Fig3S}~a) we depict the target connectivity. In Figs.~\ref{Fig3S}~b)~and~c) we show the matrix representation of the instantaneous and target Hamiltonian, respectively.

One interesting aspect of the hardcore boson regime is that we can also simulate all-to-all connectivities. In fact,  by using the same driving driving sequence $-5J<g_l<5J$ depicted in Fig. 2 l) of the main manuscript, we can simulate the dynamics of $M=1,2,3,4$ excitations in a device with a ring connectivity. 
The reason for this is that in the hardcore boson regime, the bosonic excitations are effectively fermionized. Thus, each site can host only one excitation. As they do not interact, one can obtain the solution of the manybody problem by invoking the driving protocol of a single excitation. In other words, the manybody state can be constructed by using a slater determinant involving single-particle wave functions. In Figs.~\ref{Fig4S}, we illustrate the effective Hamiltonians for $M=1,2,3,4$ hardcore bosons in a processor with all-to-all connectivity.
We envision that these results will have an impact on simulations of quantum chemistry problems.

\section{A minimal example of a 3SAT problem and its adiabatic deformation}
\label{AppendixD}

In this section, we explain the basic elements on 3SAT problems and its algebraic origin. Let us begin by considering
the system of equations given in the main text
\begin{align}
1&=a_{2}+a_{3}+a_{1}a_{3},\label{ex1:sys1:eq1}\\
1&=a_{1}+a_{3}+a_{1}a_{2},\label{ex1:sys1:eq2}\\
0&=a_{1}+a_{2}+a_{2}a_{3}.\label{ex1:sys1:eq3}
\end{align}

For each Equation (\ref{ex1:sys1:eq1}), (\ref{ex1:sys1:eq2}) and (\ref{ex1:sys1:eq3}) there is an objective function that indicates if an assignment of $(a_1,a_2,a_3)$ \emph{SAT}isfies the corresponding equation. We sum up the three objective functions into one that counts the number of solutions for the entire system.

To find an objective function (not necessarily unique), a simple recipe is to use the inclusion-and-exclusion principle. Let $C_1$, $C_2$ and $C_3$ denote respectively objective functions for Equations (\ref{ex1:sys1:eq1}), (\ref{ex1:sys1:eq2}) and (\ref{ex1:sys1:eq3}). Then we have
\begin{widetext}
\begin{align}
C_1&=1-(a_{2}+a_{3}+a_{3}a_{1}-2(a_{2}a_{3}+a_{1}a_{3}+a_{1}a_{2}a_{3})+4a_{1}a_{2}a_{3}),\\
C_2&=1-(a_{1}+a_{3}+a_{1}a_{2}-2(a_{1}a_{3}+a_{1}a_{2}+a_{1}a_{2}a_{3})+4a_{1}a_{2}a_{3}),\\
C_3&=a_{1}+a_{2}+a_{2}a_{3}-2(a_{1}a_{2}+a_{2}a_{3}+a_{1}a_{2}a_{3})+4a_{1}a_{2}a_{3}.
\end{align}
\end{widetext}
We sum the former relations and then obtain the following objective function for the entire system:
\begin{align}
C&=2-2a_{3}+3a_{1}a_{3}+a_{2}a_{3}-a_{1}a_{2}-2a_{1}a_{2}a_{3}\label{global_obj_fct}.
\end{align}
Let's check that our objective function is right.  To do this we construct the table 

\begin{center}
	\begin{tabular}{|c|c|c||c|}
		\hline
		$a_{1}$ & $a_{2}$ & $a_{3}$ & $C$\\
		\hline\hline
		$0$ & $0$ & $0$ & $2$\\
		\hline
		$0$ & $0$ & $1$ & $0$\\
		\hline
		$0$ & $1$ & $0$ & $2$\\
		\hline
		$0$ & $1$ & $1$ & $1$\\
		\hline
		$1$ & $0$ & $0$ & $2$\\
		\hline
		$1$ & $0$ & $1$ & $3$\\
		\hline
		$1$ & $1$ & $0$ & $1$\\
		\hline
		$1$ & $1$ & $1$ & $1$\\
		\hline
	\end{tabular}
\end{center}

As the original system represents a permutation, the values of $C$ must follow those of the binomial coefficients $\binom{3}{i}$ for $0\leq i\leq 3$. This is not a coincidence and can be easily seen from the inclusion-and-exclusion principle. In a similar way than in adiabatic computation, the initial Hamiltonian is usually tailored such that the algebraic multiplicities of its eigenvalues follow those of the binomial coefficient.

We want to point out that Equations (\ref{ex1:sys1:eq1}), (\ref{ex1:sys1:eq2}), and (\ref{ex1:sys1:eq3}) are over $\mathbb{Z}_{2}$ which involve addition and multiplication modulo $2$. Some readers might be acquainted with SAT problems involving \texttt{true} or \texttt{false} together with logical operations of conjunction (and), disjunction (or) and negation (not). There is a one-to-one correspondence between expressions over $\mathbb{Z}_{2}$ and expressions over $\{\texttt{true},\texttt{false}\}$ involving operations and, or and not.

For instance the two simplest non-trivial \emph{irreducible} expressions that lead to 3-SAT problems containing $3$ variables are given by
\begin{displaymath}
x+yz\quad\text{and}\quad x+y+z.
\end{displaymath}
We observe that 
\begin{align*}
x+yz&\Leftrightarrow \big(x\wedge\neg(y\wedge z)\big)\vee\big(\neg x\wedge (y\wedge z)\big)\\
&\Leftrightarrow \big(x\wedge(\neg y)\big)\vee\big(x\wedge(\neg z)\big)\vee\big((\neg x)\wedge y\wedge z\big).
\end{align*}
Similarly, we can find a logical 3SAT expression for the algebraic 3SAT expression $x+y+z$.

In our example involving Equations (\ref{ex1:sys1:eq1}), (\ref{ex1:sys1:eq2}), and (\ref{ex1:sys1:eq3}), there is no point to rewrite the expression into their logical flavour. This is because it is straightforward to obtain the objective function directly from the algebraic 3SAT as we show.

Also since there are efficient classical algorithms to solve linear systems of equations over any algebraic field (and even extremely efficient ones over $\mathbb{Z}_{2}$), there is no point to solve such system with quantum devices. Therefore this explains why we look upon $3$-variable systems that are quadratic such as the one obtained from a Toffoli permutation or such as the more cryptographic one given by Equations (\ref{ex1:sys1:eq1}), (\ref{ex1:sys1:eq2}), and (\ref{ex1:sys1:eq3}).

The example given by Equations (\ref{ex1:sys1:eq1}), (\ref{ex1:sys1:eq2}), and (\ref{ex1:sys1:eq3}) (below in Section 3) is a typical (toy) example of a system of equations that arises when performing cryptanalysis of block ciphers or hash functions. The algebraic degree of every equation is $2$, which is high with respect to the maximum possible degree that is $3$. The number of terms per equation is high with respect to the maximal number of possible terms which is $8$ in our case. Expressed differently and in an equivalent way, the density for the number of terms is relatively high. For more information on properties that matter to system of equations from cryptography, see ~\cite{GraPanThoUniPer2019}~and ~\cite{MulPanHanFin2013}. There are $8$ possible assignments to the system of equations. The solution space of the system of equations is mapped to the minimal value of the objective function.

\subsection{Obtaining Hamiltonians from cost functions to solve 3SAT instances}

Let us recall the objective function from (\ref{global_obj_fct}) which is
\begin{displaymath}
C=2-2a_{3}+3a_{1}a_{3}+a_{2}a_{3}-a_{1}a_{2}-2a_{1}a_{2}a_{3}.
\end{displaymath}
We use an affine transformation to map a variable that appears in a term from (\ref{global_obj_fct}) into a $2\times 2$ diagonal matrix:
\begin{align}
a_{1}&\mapsto \frac{1}{2}\big(1+\mu^x_{1}\big), \quad a_{2}\mapsto \frac{1}{2}\big(1+\mu^x_{2}\big),\quad a_{3}\mapsto \frac{1}{2}\big(1-\mu^x_{3}\big),
\end{align}
A product of variables is mapped to the kronecker product of the diagonal matrices. It is understood that a variable which does not appear in a product is mapped to the identity. Our choice of ordering the indices of the subsystems is $(3,2,1)$. More precisely for our cost function, we have the operator

 %%%
\begin{align}
\label{eq:3SATHamiltonianCost}
\hat{\mathcal{C}}=\frac{1}{4}(\mu^x_{1}-\mu^x_{2}+\mu^x_{3}-2\mu^x_{1}\mu^x_{2}-2\mu^x_{1}\mu^x_{3}+\mu^x_{1}\mu^x_{2}\mu^x_{3})
 \end{align}
%%%
associated to the cost function $C(a_1,a_2,a_3)$.

%%%%%%%%
\begin{figure}
	\centering
	\includegraphics[width=0.50\textwidth]{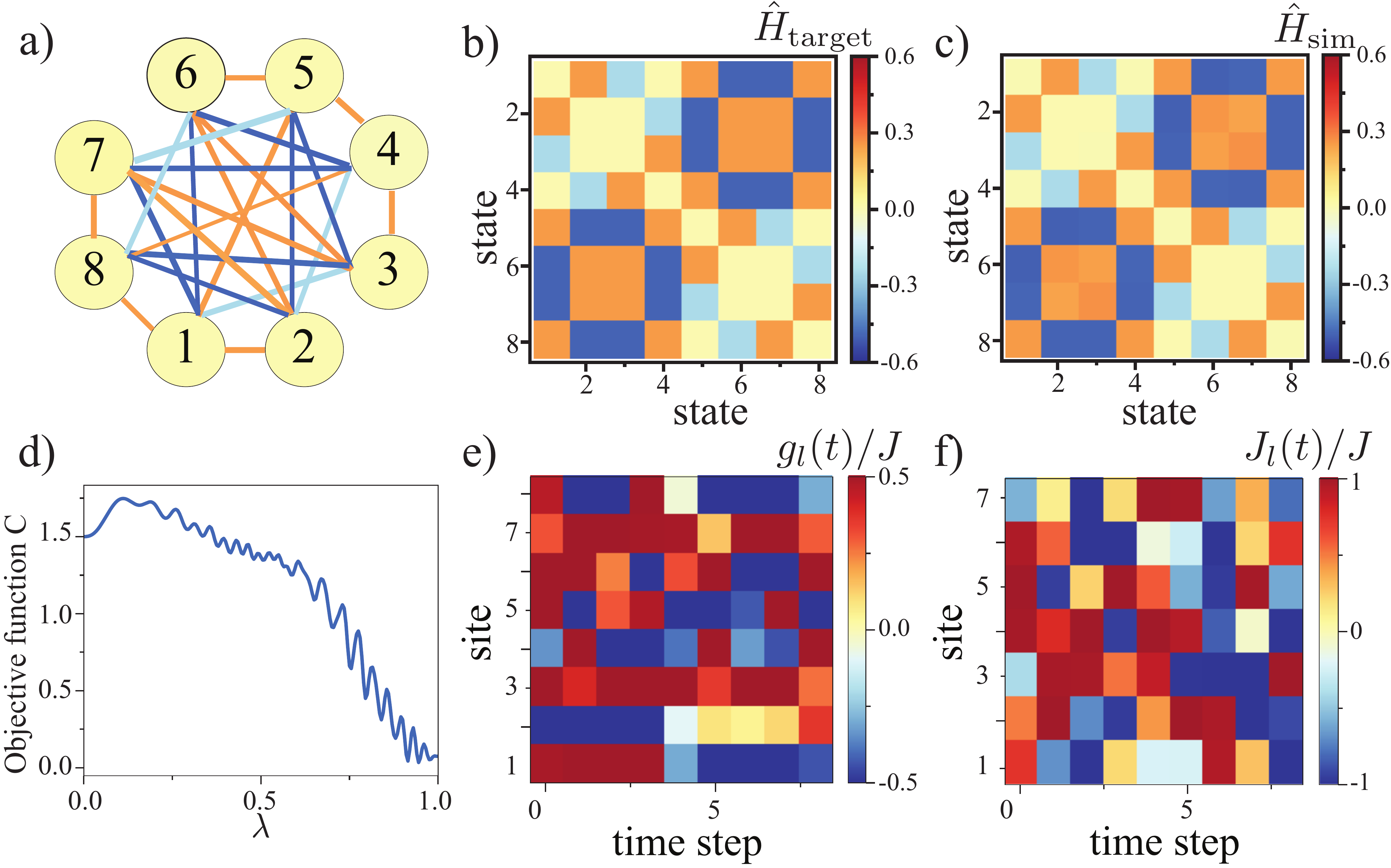}
	\caption{Adiabatic deformation of the 3SAT Hamiltonian. a) Illustrates the target connectivity that we want to simulate with a single excitation $M=1$ in $L=8$ sites. b) Depicts the matrix representation of the 3SAT Hamiltonian (target Hamiltonian) in the basis $\ket{1_l}=\ket{0,0,\ldots,1_l,0,0}$. c) Shows the solution obtained using GRAPE for effective Hamiltonian that we can stroboscopically simulate. In d) we depict the expectation value of the cost function $\hat{\mathcal{C}}=\frac{1}{4}(\mu^x_{1}-\mu^x_{2}+\mu^x_{3}-2\mu^x_{1}\mu^x_{2}-2\mu^x_{1}\mu^x_{3}+\mu^x_{1}\mu^x_{2}\mu^x_{3})$ in the instantaneous ground state of the effective Hamiltonian. In e)  and  f) we also illustrate the driving protocol of the onsite energies $g_l(t)$ and the couplings $J_l(t)$ obtained using GRAPE. The period of the driving sequence is set to be $T=6.38/J$ and the driving sequence $-5J<g_l<5J$, $-J<J_l<J$ is divided into $N=11$ time steps. The adiabatic evolution happens over 200 cycles, makes a total time of $T_\text{total}=1276/J$.
}
	\label{Fig5S}
\end{figure}
%%%%%%%%

In our manuscript, we scale this operator and define the Hamiltonian

 %%%
\begin{align}
\label{eq:3SATHamiltonianApp}
\hat{H}_{\text{SAT}}&=\hbar\omega(\mu^x_{1}-\mu^x_{2}+\mu^x_{3}-2\mu^x_{1}\mu^x_{2}-2\mu^x_{1}\mu^x_{3}+\mu^x_{1}\mu^x_{2}\mu^x_{3})
\ ,
 \end{align}
%%%
where $\omega$ has units of angular frequency. Once we have the Hamiltonian, the solution of the 3SAT problem is encoded in the ground state $\ket{G}$. In order to obtain to the triplet $(a_1, a_2, a_3)$ satisfying  the 3 clauses discussed above, we just need to calculate the following expectation values
\begin{widetext}
%%%
\begin{align}
\label{eq:Formulas}
a_1&=\frac{1}{2}\bra{G}\left[1+\frac{1}{2}(\sigma^x_1 \sigma^x_5+ \sigma^y_1 \sigma^y_5 + \sigma^x_2 \sigma^x_6+ \sigma^y_2 \sigma^y_6 +  \sigma^x_3 \sigma^x_7+ \sigma^y_3 \sigma^y_7 + \sigma^x_4 \sigma^x_8+ \sigma^y_4 \sigma^y_8  )\right]\ket{G}
\nonumber\\
a_2&=\frac{1}{2}\bra{G}\left[1+\frac{1}{2}( \sigma^x_1 \sigma^x_3 + \sigma^y_1 \sigma^y_3 +  \sigma^x_2 \sigma^x_4 + \sigma^y_2 \sigma^y_4  +  \sigma^x_5 \sigma^x_7 + \sigma^y_5 \sigma^y_7 +  \sigma^x_6 \sigma^x_8 + \sigma^y_6 \sigma^y_8) \right]\ket{G}
\nonumber\\
a_3&=\frac{1}{2}\bra{G}\left[1-\frac{1}{2}(\sigma^x_1 \sigma^x_2+ \sigma^y_1 \sigma^y_2 +  \sigma^x_3 \sigma^x_4+ \sigma^y_3 \sigma^y_4 +  \sigma^x_5 \sigma^x_6+ \sigma^y_5 \sigma^y_6 +  \sigma^x_7 \sigma^x_8+ \sigma^y_7 \sigma^y_8) \right]\ket{G}
\ ,
 \end{align}
 %%%%
 \end{widetext}
which correspond to two-point correlations in the original basis of qubits.

%%%
\subsection{Instantaneous Hamiltonian and driving protocols for the adiabatic deformation of the 3SAT Hamiltonian}
\label{AppendixE}
%%%%

In this section we provide additional information on the adiabatic deformation of the 3SAT Hamiltonian Eq.~\eqref{eq:3SATHamiltonian}. With this aim, we consider a single excitation  $M=1$ in an array of $L=8$ sites and the dimension of the Hilbert space is $D_{1,9}=9$. As we discussed above, we consider the basis $\ket{1_l}=\ket{0,0,\ldots,1_l,0,\dots,0}$ with $l=1,\dots,L$, where $\ket{1_l}$ denotes an excitation at the $l$-th site. In this case, we are not only able to simulate the target Hamiltonian, but we also perform an adiabatic modulation of the parameters to interpolate two effective Hamiltonians. The adiabatic evolution happens over 200 cycles, makes a total time of $T_\text{total}=1276/J$. Fig.~\ref{Fig5S} a) shows the topology of the target Hamiltonian for 3-SAT problem. Figs.~\ref{Fig5S}~b)~and~c) illustrate the target Hamiltonian and the Hamiltonian obtained by using GRAPE, respectively. Fig.~\ref{Fig5S} d) Depicts the expectation value of the objective function Eq.~\eqref{eq:3SATHamiltonianCost} during the adiabatic deformation of the 3SAT Hamiltonian. In Figs.~\ref{Fig5S} ~e)~and~f) we show the driving protocol obtained using GRAPE to simulate the target Hamiltonian.

\section{Performance of our method and its dependence on different parameters}
\label{AppendixE}
%%%%%%%%
\begin{figure}
	\centering
	\includegraphics[width=0.50\textwidth]{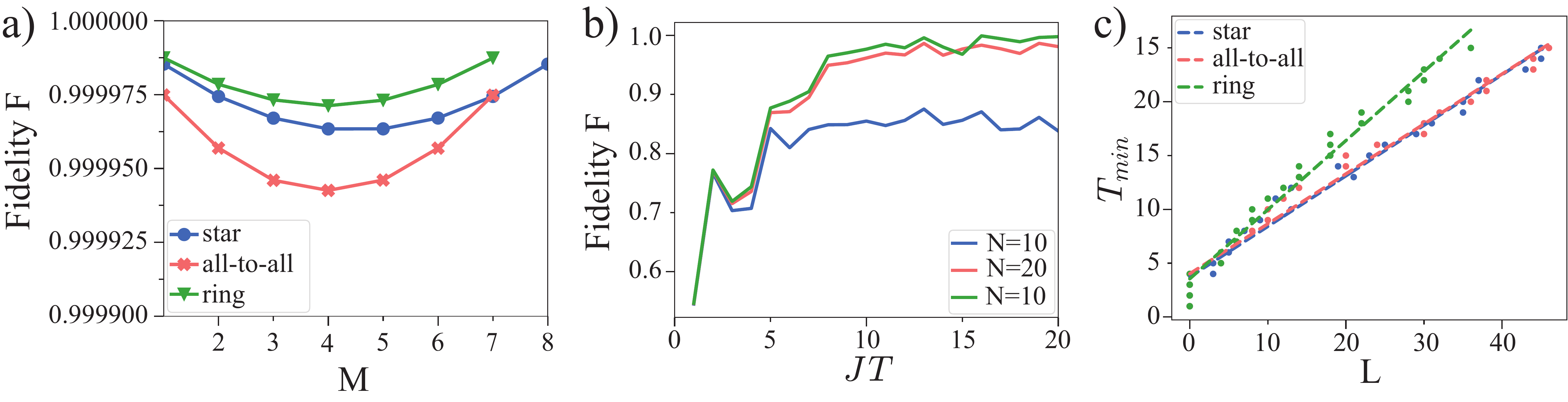}
	\caption{a) Fidelity of effective dynamics for hard-core bosons in a driven chain. All parameters are scaled in units of the nearest-neighbor coupling strength $J$. a) Fidelity for star-graph ($L=9$ sites), all-to-all coupling ($L=8$ sites) and ring ($L=8$) sites. Driving with $N=10$ time steps and time $T=10/J$, with driving of local potential in range $-5J<g_l<5J$. b) Fidelity of effective dynamics for driving time and timesteps of a linear chain with a $N_\text{p}=2$ excitation and $U=4J$. All parameters are scaled in units of the nearest-neighbor coupling strength $J$. Target Hamiltonian is all-to-all coupling  with $L=8$ sites with driving of both potential ($-5J<g_l<5J$) and nearest-neighbor coupling $-J<J_l<J$. c) Minimal time needed $T_\text{min}$ to generate effective Hamiltonian with fidelity $F>0.999$, for varying system size of chain $L$. The result is fitted with a linear equation $T_\text{min}=aL+b$ (dashed lines). The found slope is $a_\text{star}=0.47/J$, $a_\text{all}=0.46/J$ and $a_\text{ring}=0.64/J$.   All parameters are scaled in units of the nearest-neighbor coupling strength $J$. Driving of potential is bounded between $-5J<g_l<5J$ and a single time step of the driving protocol is fixed to $\tau =1/J$.
	}
	\label{Fig6S}
\end{figure}
In our manuscript, we present minimal examples of the applicability of our method. We were able to simulate  such as star, all-to-all and ring connectivities and consider the case of interacting and Hardcore bosons as well. However, one might ask what is the scalability of method and what is its dependence on other parameters such as the number of steps in the driving protocol. This is precisely the goal of this section. First we discuss the scaling of the effective Hamiltonian with the number $M$ of excitations in the hardcore boson regime. After that, we investigate scaling of the effective Hamiltonian in the case of two excitation with a finite interaction strength $U=4J$ as a function of the period $T$ and the number of steps $N$. Finally, we concentrate on the dependence of the effective Hamiltonian on the driving parameters.

\subsection{Scaling of the method with respect to different parameters of the system}
We discuss the case of many particles for the hard-core bosonic chain. Driving a hard-core chain generates arbitrary non-interacting fermionic many-body Hamiltonian.
By using GRAPE, we calculate the best driving parameters for the single particle case first. Then, the same driving is used for the same system, but now with multiple excitations $M>1$. The resulting effective Hamiltonian is then the corresponding fermionic many-body Hamiltonian. The scaling of the fidelity for star, all-to-all  and ring connectivies is shown in Fig.\ref{Fig6S}~a). In Fig.~\ref{Fig4S}, we have depicted the effective Hamiltonian for all-to-all connectivity  for different excitations.

As  we discussed above, in the hardcore boson regime, one can solve the manybody problem just by obtaining the solution for a single particle, because effectively the system is non-interacting and one can map it to a system of free fermions. However, for a finite value of the interaction $U=4J$, the excitations are far from the hardcore boson regime, and we cannot reconstruct the solution to the two body problem by investigating the single particle case.  For the Bose-Hubbard model with $M=2$ particles, we note that more time steps $N$ are needed to generate the effective Hamiltonian compared to the single particle case. To investigate this issue in detail, here we consider the case of a all-to-all connectivity and calculate the fidelity of the numerically obtained unitary operator with respect to target unitary. We explore the dependence of this fidelity as a function of the period $T$ of the driving and the number of time steps, as we depict in Fig.\ref{Fig6S}~b).

We investigate the scaling of the effective Hamiltonian generation for varying system size. The result is shown in Fig.\ref{Fig6S}~c). We show the star-graph, all-to-all coupling and ring for single excitation and potential driving. We vary the system size $L$ of the chain and calculate the minimal time needed $T_\text{min}$ to generate the effective Hamiltonian for the given configuration with a fidelity $F>0.999$. We fix the length of a time step of the driving protocol to $\tau=1/J$, where $J$ is the coupling strength of the chain. We observe an approximately linear scaling between protocol time $T_\text{min}$ and system size $L$.

\subsection{Effective Hamiltonian dependence on driving parameters}
\begin{figure}
	\centering
	\includegraphics[width=0.50\textwidth]{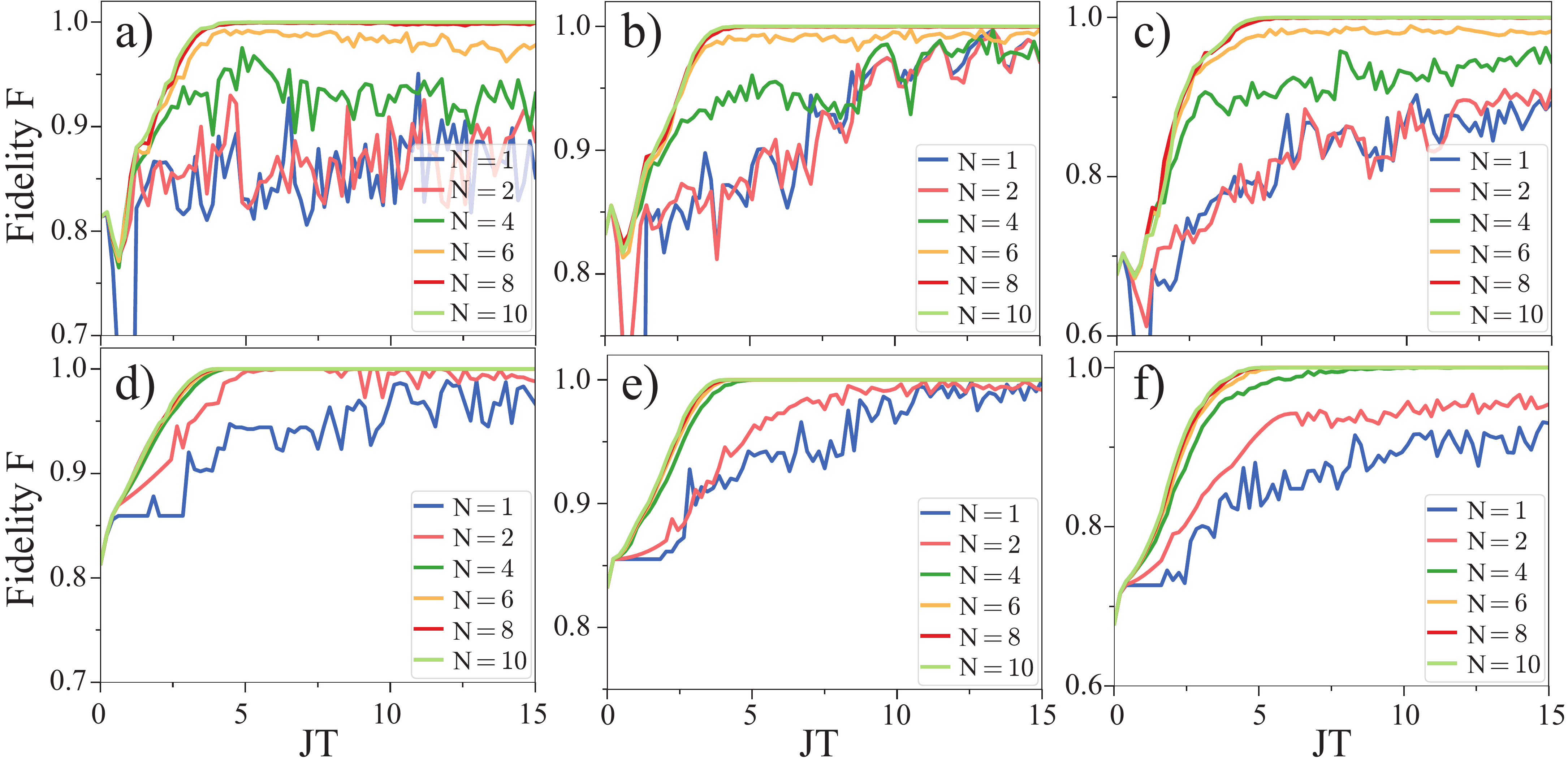}
	\caption{Fidelity of effective dynamics for driving time and time steps of a linear chain with a single excitation. All parameters are scaled in units of the nearest-neighbor coupling strength $J$. a), d) target effective Hamiltonian is the star graph with $L=9$ sites. b), e) target is all-to-all coupling  with $L=8$ sites. c),f) target is the simulation of boolean equations  with $L=8$ sites. a),b), c) Driving of local potential only in range $-5J<g_l<5J$. d,e,f) Driving of both potential ($-5J<g_l<5J$) and nearest-neighbor coupling $-J<J_l<J$.
	}
	\label{Fig7S}
\end{figure}

The fidelity of the effective Hamiltonian that can be generated depends on several parameters. We investigate here how the fidelity of the effective dynamics is affected by  the driving time $T$ as well as the number $N$ of discrete steps of the driving protocol. The results for a driven linear chain are presented in Fig.\ref{Fig7S}. We generate a star graph [Figs.\ref{Fig7S}~a),~d)], all-to-all coupling [Figs.\ref{Fig7S}~b),~e)] and the boolean equations [Figs.\ref{Fig7S}~c),~f)]. We use either driving of local potential only [Figs.\ref{Fig7S}~a),~b),~c)] or drive both potential and nearest-neighbor couplings [Figs.\ref{Fig7S}~d),~e),~f)]. 
We observe that when only the local potential is driven, at least $N=8$ steps are needed to achieve sufficient fidelity. For driving both potential and nearest-neighbor coupling, 4 time steps are sufficient. There is also a minimal time needed before maximal fidelity is reached, which is nearly the same for all three problems and on the order of $T_\text{min}=4/J$.

%%%%%%%%
%%%%%%%%
\section{Explicit matrix representation of the target Hamiltonian of the LiH molecule}
\label{AppendixF}
The corresponding Hamiltonian matrix for LiH is as follows
\begin{widetext}
\setcounter{MaxMatrixCols}{20}
\begingroup
\footnotesize
\begin{equation*}
\begin{pmatrix}
.00846 & -.33392 & .03370 & -.21996 & .33392 & -.08335 & .10560 & -.08001 & -.03370 & .10560 & -.18833 & .09292 & -.21996 & .08001 & -.09292 & .10347 \\
-.33392 & -.02541 & -.21713 & .07039 & -.08335 & .37960 & -.04784 & .09209 & .10560 & -.17922 & -.00038 & -.13611 & .08001 & -.26748 & .05141 & -.09746 \\
.03370 & -.21713 & -.00029 & -.34549 & .10560 & -.04784 & .35689 & -.09255 & -.18833 & -.00038 & -.10054 & .13207 & -.09292 & .05141 & -.25612 & .09590 \\
-.21996 & .07039 & -.34549 & -.03981 & -.08001 & .09209 & -.09255 & .33823 & .09292 & -.13611 & .13207 & -.05945 & .10347 & -.09746 & .09590 & -.19952 \\
.33392 & -.08335 & .10560 & -.08001 & -.02541 & -.37960 & .17922 & -.26748 & -.21713 & .04784 & .00038 & .05141 & -.07039 & .09209 & -.13611 & .09746 \\
-.08335 & .37960 & -.04784 & .09209 & -.37960 & .15372 & -.11896 & .14861 & .04784 & -.11896 & .11505 & -.03415 & .09209 & -.14861 & .03415 & -.12120 \\
.10560 & -.04784 & .35689 & -.09255 & .17922 & -.11896 & .07367 & -.42903 & .00038 & .11505 & -.17038 & .03095 & -.13611 & .03415 & -.11232 & .11348 \\
-.08001 & .09209 & -.09255 & .33823 & -.26748 & .14861 & -.42903 & -.04424 & .05141 & -.03415 & .03095 & -.19948 & .09746 & -.12120 & .11348 & -.06392 \\
-.03370 & .10560 & -.18833 & .09292 & -.21713 & .04784 & .00038 & .05141 & -.00029 & -.35689 & .10054 & -.25612 & .34549 & -.09255 & .13207 & -.09590 \\
.10560 & -.17922 & -.00038 & -.13611 & .04784 & -.11896 & .11505 & -.03415 & -.35689 & .07367 & -.17038 & .11232 & -.09255 & .42903 & -.03095 & .11348 \\
-.18833 & -.00038 & -.10054 & .13207 & .00038 & .11505 & -.17038 & .03095 & .10054 & -.17038 & .05177 & -.39058 & .13207 & -.03095 & .39058 & -.11241 \\
.09292 & -.13611 & .13207 & -.05945 & .05141 & -.03415 & .03095 & -.19948 & -.25612 & .11232 & -.39058 & -.04576 & -.09590 & .11348 & -.11241 & .34422 \\
-.21996 & .08001 & -.09292 & .10347 & -.07039 & .09209 & -.13611 & .09746 & .34549 & -.09255 & .13207 & -.09590 & -.03981 & -.33823 & .05945 & -.19952 \\
.08001 & -.26748 & .05141 & -.09746 & .09209 & -.14861 & .03415 & -.12120 & -.09255 & .42903 & -.03095 & .11348 & -.33823 & -.04424 & -.19948 & .06392 \\
-.09292 & .05141 & -.25612 & .09590 & -.13611 & .03415 & -.11232 & .11348 & .13207 & -.03095 & .39058 & -.11241 & .05945 & -.19948 & -.04576 & -.34422 \\
.10347 & -.09746 & .09590 & -.19952 & .09746 & -.12120 & .11348 & -.06392 & -.09590 & .11348 & -.11241 & .34422 & -.19952 & .06392 & -.34422 & -.05026 \\
\end{pmatrix}
\end{equation*}
\endgroup
\end{widetext}

%\bibliographystyle{phaip}

%\bibliography{library,XbibTobias.bib,MBLTobias,MybibTobias}
%\bibliographystyle{unsrt}
%\bibliography{combined}


\end{document}